\documentclass[prd,superscriptaddress,twocolumn,showpacs,amsmath,%
preprintnumbers]{revtex4}
\usepackage{eucal} 
\usepackage{bm} 
\usepackage{graphicx}
\newcommand{\beq}{\begin{equation}}
\newcommand{\eeq}{\end{equation}}
\newcommand{\be}{\begin{eqnarray}}
\newcommand{\ee}{\end{eqnarray}}
\begin{document}
\title{Chiral dynamics and partonic structure 
at large transverse distances}
\author{M.~Strikman}
\affiliation{Department of Physics, Pennsylvania State University,
University Park, PA 16802, USA}
\author{C.~Weiss}
\affiliation{Theory Center, Jefferson Lab, Newport News, VA 23606, USA}
\begin{abstract}
We study large--distance contributions to the nucleon's parton densities
in the transverse coordinate 
(impact parameter) representation based on generalized parton distributions 
(GPDs). Chiral dynamics generates a distinct component of the partonic 
structure, located at momentum fractions $x \lesssim M_\pi / M_N$ and 
transverse distances $b \sim 1/M_\pi$. We calculate this component using 
phenomenological pion exchange with a physical lower limit in $b$ 
(the transverse ``core'' radius estimated from the nucleon's 
axial form factor, $R_{\rm core} = 0.55 \, \textrm{fm}$) and demonstrate 
its universal character. This formulation preserves the basic picture 
of the ``pion cloud'' model of the nucleon's sea quark distributions, 
while restricting its application to the region actually governed
by chiral dynamics. It is found that (a) the large--distance component 
accounts for only $\sim 1/3$ of the measured antiquark flavor asymmetry 
$\bar d - \bar u$ at $x \sim 0.1$;
(b) the strange sea quarks, $s$ and $\bar s$, are significantly more 
localized than the light antiquark sea; (c) the nucleon's singlet quark 
size for $x < 0.1$ is larger than its gluonic size, 
$\langle b^2 \rangle_{q + \bar q} > \langle b^2 \rangle_{g}$,
as suggested by the $t$--slopes of deeply--virtual Compton scattering 
and exclusive $J/\psi$ production measured at HERA and FNAL. 
We show that our approach reproduces the general $N_c$--scaling 
of parton densities in QCD, thanks to the degeneracy of 
$N$ and $\Delta$ intermediate states in the large--$N_c$ limit.
We also comment on the role of pionic configurations
at large longitudinal distances and the limits of their 
applicability at small $x$.
\end{abstract}
%
%
%\keywords{Chiral dynamics, parton distributions, pion cloud model, 
%hard exclusive processes}
%
%
\preprint{JLAB-THY-09-1012}
\maketitle
\tableofcontents
\section{Introduction}
Parton densities summarize the structure of the nucleon probed
in high--momentum transfer processes such as deep--inelastic 
lepton--nucleon scattering and production of high--mass systems
(jets, heavy particles) in nucleon--nucleon collisions. They are 
defined in the context of a factorization procedure, by which the 
cross section of these processes is separated into a short--distance 
quark/gluon subprocess, calculable in perturbative QCD, and the 
distribution of the partons in the initial state, and thus represent 
long--distance, low--energy characteristics of the nucleon. 
As such, they are governed by the same low--energy dynamics which 
determines other nucleon observables like the vector and axial couplings 
(to which they are related by the partonic sum rules), form factors, 
meson--nucleon couplings \textit{etc.} Of particular interest are 
the charge ($u - \bar u, d - \bar d$) and flavor ($u - d, \bar u - \bar d$)
non--singlet quark densities, which exhibit 
only weak scale dependence and are of non--perturbative origin; 
they represent quasi--observables which directly probes the QCD 
quark structure of the nucleon at low resolution scales. 

The long--distance behavior of strong interactions at low energies is 
governed by the spontaneous breaking of chiral symmetry in QCD. 
The Goldstone boson nature of the pion explains its small mass 
on the hadronic scale and requires its coupling to other
hadrons to vanish in the long--wavelength limit. The resulting
``chiral dynamics'' gives rise to a number of distinctive phenomena at
distance scales $\sim 1/M_\pi$, such as the $\pi\pi, \pi N$ and $NN$ 
interaction at large distances, the pion pole in the axial current
matrix element, \textit{etc.} An important question is how chiral 
dynamics affects the nucleon's parton densities, and whether one can
see any signs of chiral effects in observables of high--momentum 
transfer processes.

The prime candidate for an effect of chiral dynamics in parton densities 
has been the flavor asymmetry of the light antiquark densities in 
the nucleon. Measurements of the proton--neutron structure function 
difference in inclusive deep--inelastic scattering \cite{Amaudruz:1991at},
semi--inclusive meson production \cite{Ackerstaff:1998sr}, 
and particularly Drell--Yan pair 
production \cite{Baldit:1994jk,Hawker:1998ty,Towell:2001nh} 
have unambiguously shown that $\left[\bar d - \bar u\right] (x) > 0$ 
in the proton for $x < 0.3$, and have partly mapped the 
$x$--dependence of the asymmetry; see Refs.~\cite{Kumano:1997cy}
for a review of earlier experimental results. The basic picture
is that the ``bare'' proton can make a transition to a virtual state 
containing a pion, and fluctuations $p \rightarrow n \pi^+$ are more likely 
than $p \rightarrow \Delta^{++} \pi^-$, resulting in an excess of $\pi^+$
over $\pi^-$ in the ``dressed'' proton. Following the original prediction
of Ref.~\cite{Thomas:1983fh}, which included only the nucleon intermediate 
state, this idea was implemented in a variety of dynamical models, 
which incorporate finite--size effects through various types of hadronic 
form factors associated with the $\pi N N$ and $\pi N \Delta$ vertices; 
see Refs.~\cite{Kumano:1997cy} for a review of the extensive literature. 
It was noted long ago \cite{FMS} that in order to reproduce the fast 
decrease of the observed asymmetry with $x$ one needs $\pi NN$ form 
factors much softer than those commonly used in meson exchange 
parametrizations of the $NN$ interaction \cite{Machleidt:hj}. 
However, even with such soft form factors the pion transverse momenta 
in the nucleon generally extend up to values $\gg M_\pi$ \cite{Koepf:1995yh}. 
This raises the question to what extent such models actually describe 
long--distance effects associated with soft pion exchange 
(momenta $\sim M_\pi$), and what part of their predictions
is simply a parametrization of short--distance dynamics which 
should more naturally be described in terms of non-hadronic degrees of
freedom. More generally, one faces the question how to formulate the
concept of the ``pion cloud'' in the nucleon's partonic structure
\cite{Sullivan:1971kd} in a manner consistent with chiral dynamics in QCD.

A framework which allows one to address these questions in a 
systematic fashion is the transverse coordinate (or impact parameter)
representation, in which the distribution of partons is studied 
as a function of the longitudinal momentum fraction, $x$, and
the transverse distance, $b$, of the parton from the transverse 
center--of--momentum of the nucleon \cite{Burkardt:2002hr,Pobylitsa:2002iu}. 
In this representation, chiral dynamics can be associated with 
a distinct component of the partonic structure, located at 
$x \lesssim M_\pi / M_N$ and $b \sim 1/M_\pi$.
In a previous work \cite{Strikman:2003gz} we have shown that in the 
gluon density this large--distance component is sizable and causes 
the nucleon's average gluonic transverse size, 
$\langle b^2 \rangle_g$ to grow if $x$ drops below $M_\pi / M_N$, 
in agreement with the $t$--slopes observed in exclusive
$J/\psi$ photo-- and electroproduction at 
HERA \cite{Aktas:2005xu,Chekanov:2004mw}, FNAL \cite{Binkley:1981kv}, 
and experiments at lower energies. 
Essential in this is the fact that in the gluon density 
(more generally, in any isoscalar parton density)
the pion cloud contributions from $N$ and $\Delta$ intermediate 
states have the same sign and add constructively. The special role of 
the $\Delta$ compared to other excited baryon states is supported 
by the fact that in the large--$N_c$ limit of QCD the $N$ and $\Delta$ 
are degenerate and enter on an equal footing.

In this article we perform a comprehensive study of the chiral 
large--distance component of the nucleon's partonic structure,
considering both its contribution to the total quark/antiquark/gluon 
densities and to the nucleon's average partonic transverse size.
The method we use to calculate this component is phenomenological 
pion exchange formulated in the impact parameter representation, 
restricted to the region of large transverse distances. A physical lower 
limit in $b$ for $\pi B \; (B = N, \Delta)$ configurations in the nucleon
wave function is set by the transverse ``core'' radius, 
estimated from the nucleon's axial form factor, 
$R_{\rm core} = 0.55 \, \textrm{fm}$, 
and we explicitly demonstrate the universal character of the 
pionic contributions in the region $b > R_{\rm core}$. This formulation 
preserves the basic physical picture of the ``pion cloud'' model of the 
nucleon's sea quark distributions, while restricting its application 
to the region actually governed by chiral dynamics. 
In fact, our study serves both a conceptual and a practical purpose. 
First, we want to establish in which region of transverse distances 
the results of the traditional pion cloud model are model--independent
and can be associated with large--distance chiral dynamics. 
Second, we want to employ this model to actually calculate the
universal large--distance component and study its properties. 
A preliminary account of our study of the flavor asymmetry 
$\bar d - \bar u$ was presented in Ref.~\cite{Strikman:2008wb}.

The investigation reported here proceeds in several steps. 
In Sec.~\ref{sec:chiral}, we develop the theory of large--distance
contributions to the partonic structure from a general, 
model--independent perspective. We outline the parametric
region of $\pi B$ configurations in the nucleon wave function,
the properties of the $b$--dependent momentum distribution 
of pions in the nucleon, its large--$b$ asymptotics, 
and the convolution formulas for the nucleon parton densities. 
In Sec.~\ref{sec:model}, we investigate the phenomenological
pion cloud model in the impact parameter representation, 
and demonstrate that at large $b$ its predictions become independent 
of the $\pi N B$ form factors modeling the short--distance dynamics. 
We also comment on the extension
of this model to $SU(3)$ flavor. In Sec.~\ref{sec:decomposition}
we then apply this model to calculate the large--distance contributions
to the sea quark distributions in the nucleon, including the isovector
$(\bar d - \bar u)$ and isoscalar $(\bar u + \bar d)$ light quark
sea, the strange sea ($s, \bar s$), and the $SU(3)$--flavor symmetry 
breaking asymmetry $(\bar u + \bar d - 2\bar s)$. We compare the
calculated large--distance contributions to empirical parametrizations 
of the parton densities and thus indirectly infer the contribution
from the short--distance region (``core''), which cannot be
calculated in a model--independent way. In the course of this we see
how the restriction to large $b$ solves several problems inherent in 
the traditional pion cloud model which formally allows for pionic 
configurations also at small impact parameters. 
In Sec.~\ref{sec:size} we consider the
large--distance contributions to the nucleon's partonic transverse 
size $\langle b^2 \rangle$, which is accessible experimentally through
the $t$--slope of hard exclusive processes $\gamma^\ast N 
\rightarrow M + N \, (M = \text{meson}, \gamma, \textit{etc.})$.
Because of the emphasis on large distances this quantity is 
calculable in a practically model--independent manner and 
represents a clean probe of chiral dynamics in the partonic structure.
Specifically, we show that at $x \sim 10^{-2}$ the large--distance 
contribution to the nucleon's singlet quark transverse size, 
$\langle b^2 \rangle_{q + \bar q}$, is larger than that to the
gluonic size, $\langle b^2 \rangle_{g}$, which is consistent with 
the observation of a larger $t$--slope in deeply--virtual Compton
scattering \cite{Aaron:2007cz,Chekanov:2008vy} 
than in exclusive $J/\psi$ production at 
HERA \cite{Aktas:2005xu,Chekanov:2004mw}.
In Sec.~\ref{sec:largenc} we discuss the correspondence
of the phenomenological pion exchange contribution to the nucleon
parton densities with the large--$N_c$ limit of QCD. In particular,
we show that the large--distance contributions obtained from pion 
exchange reproduce the general $N_c$--scaling of parton densities in QCD, 
thanks to the degeneracy of $N$ and $\Delta$ intermediate states in the 
large--$N_c$ limit. This re-affirms the need to include intermediate
$\Delta$ states on the same footing as the nucleon, and shows that the 
phenomenological large--distance contributions considered here are
a legitimate part of the nucleons partonic structure in large--$N_c$ QCD.
Finally, in Sec.~\ref{sec:smallx} we focus on the physical limitations 
to the picture of individual $\pi B$ configurations at small $x$, arising
from the non--chiral growth of the transverse sizes due to diffusion,
and from chiral corrections to the structure of the pion. We also
comment on the role of chiral dynamics at large longitudinal distances.
Our summary and outlook are presented in Sec.~\ref{sec:summary}.
The two appendices present technical material related to the
meson--nucleon coupling constants for $SU(3)$ flavor symmetry,
and the numerical evaluation of the $b$--dependent pion 
momentum distributions in the nucleon.
 
In the context of our studies of the strange sea quark distributions, 
$s(x)$ and $\bar s(x)$, and the $SU(3)$ flavor symmetry--breaking 
asymmetry, $\left[ \bar u + \bar d - 2\bar s\right] (x)$, we consider 
also contributions from configurations containing $SU(3)$ octet mesons 
($K\Lambda, K\Sigma, K\Sigma^\ast, \eta N$) to the nucleon's partonic 
structure at large distances. 
While such high--mass configurations are not governed by chiral dynamics 
and treated at a purely phenomenological level, it is interesting to 
compare their large--distance tails with those of chiral contributions 
from pions. We note that the issue of the strange sea in the nucleon 
($s, \bar s$) and the question of possibly different $x$--distributions of
$s$ and $\bar s$ has acquired new urgency following the results of the
NuTeV experiment in semi--inclusive charged--current neutrino DIS,
which can discriminate between $s$ and $\bar s$ via the process
$W^+ + s \rightarrow c$ \cite{Goncharov:2001qe,Mason:2007zz}.

Chiral contributions to the nucleon's parton densities have been 
studied extensively within chiral perturbation theory 
\cite{Chen:2001eg,Arndt:2001ye}, mostly with the aim of 
extrapolating lattice QCD results obtained at large pion
masses toward lower values \cite{Detmold:2001jb}. Chiral perturbation theory 
was also applied to GPDs, including the impact parameter representation
\cite{Belitsky:2002jp,Ando:2006sk,Diehl:2006ya,Kivel:2002ia}. 
Compared to these calculations, which use methods of 
effective field theory based on a power--counting scheme, we take
here a more pragmatic approach. We study the pion distribution in
the nucleon in a phenomenological approach which incorporates the
finite bare nucleon size through form factors, and investigate numerically 
in which region the results become insensitive to the form factors
and can be attributed to universal chiral dynamics
\footnote{The approach taken here bears some similarity to the
use of finite--size regulators in the chiral extrapolation of
lattice QCD results \cite{Young:2005tr}.}. In this approach 
we maintain exact relativistic kinematics (physical pion and nucleon masses) 
and calculate distributions of finite support, which are then analyzed 
in the different parametric regions and matched with the asymptotic
``chiral'' predictions. This also allows us to deal with the strong
cancellations between contributions from $N$ and $\Delta$ intermediate 
states in the isovector quark densities, which are difficult to 
accommodate within a power counting scheme. In fact, the cancellation 
becomes exact in the large--$N_c$ limit of QCD and ensures 
the proper $1/N_c$ counting required of the isovector antiquark 
distribution in QCD \cite{Strikman:2003gz}.

In this study we focus on chiral large--distance contributions to the
nucleon's partonic structure at moderately small momentum fractions,
$x \gtrsim 10^{-2}$, which arise from individual $\pi B \, (B = N, \Delta)$ 
configurations in the nucleon wave function. When extending the 
discussion toward smaller $x$, several effects need to be taken into 
account which potentially modify this picture. One is diffusion in the
partonic wave function, which causes the transverse size of the nucleon's 
partonic configurations to grow at small $x$ (however, this effect
is suppressed at large $Q^2$). Another effect are possible chiral 
corrections to the structure of the pion itself, which were recently 
studied in an approach based on resummation of chiral perturbation theory 
in the leading logarithmic approximation \cite{Kivel:2008ry}. 
We discuss the limitations to the applicability of the picture of 
individual $\pi B$ configurations in Secs.~\ref{subsec:diffusion} 
and \ref{subsec:chiral_pion}. We also comment on the role of
$\pi B$ configurations at large longitudinal separations and arbitrary
transverse distances, and point out that there may be a window for a 
chiral regime at $x \gtrsim \sim 10^{-2}$; at smaller $x$ coherence
effects become dominant; see Sec.~\ref{subsec:longitudinal}. A detailed 
investigation of this new regime will be the subject of a separate study.
\section{Chiral dynamics and partonic structure}
\label{sec:chiral}
\subsection{Parametric region of chiral component}
\label{subsec:parametric}
As the first step of our study we want to delineate the parametric 
region where parton densities are governed by chiral 
dynamics and establish its numerical limits, as imposed by other,
non--chiral physical scales. The primary object of our discussion 
is the pion longitudinal momentum and transverse coordinate distribution 
in a fast--moving nucleon, $f_\pi (y, b)$, where $y$ is the pion momentum 
fraction. Here we introduce this concept heuristically, appealing to
its obvious physical meaning; its precise definition in terms of GPDs
and its region of applicability will be elaborated in the following.

\begin{figure}
\includegraphics[width=.32\textwidth]{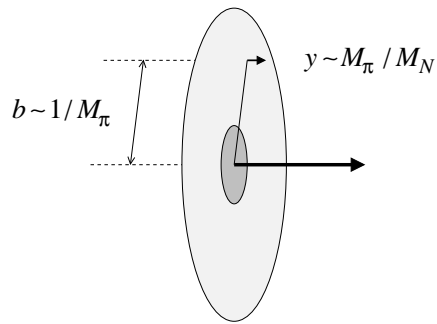}
\caption[]{Parametric region where the pion distribution in the nucleon
is governed by chiral dynamics. The variables are the pion longitudinal 
momentum fraction, $y$, and transverse position, $b$.} 
\label{fig:chiral}
\end{figure}

Chiral dynamics generally governs contributions to nucleon observables 
from large distances, of the order $1/M_\pi$, which is assumed here 
to be much larger than all other hadronic length scales in question.
These contributions result from exchange of ``soft'' pions in the 
nucleon rest frame; in the time--ordered formulation of relativistic
dynamics these are pions with energies $E_\pi \sim M_\pi$ and momenta 
$|\bm{k}_\pi| \sim M_\pi$. Chiral symmetry provides that such pions 
couple weakly to the nucleon and to each other, so that their effects 
can be computed perturbatively. Boosting these weakly interacting 
pion--nucleon configurations to a frame in which the nucleon is moving 
with large velocity, we find that they correspond to longitudinal 
pion momentum fractions of the order \footnote{In invariant perturbation 
theory soft pions
have virtualities $-k_\pi^2 \sim M_\pi^2$, and Eq.~(\ref{y_chiral}) 
results from the condition that the minimum pion virtuality which is 
kinematically required for a given longitudinal momentum fraction, $y$, 
be at most of the order $M_\pi^2$; \textit{cf}.\ Eq.~(\ref{t_min}) below.} 
\begin{equation}
y \;\; \sim \;\; M_\pi / M_N .
\label{y_chiral}
\end{equation}
At the same time, the soft pions' transverse momenta, which are not 
affected by the boost, correspond to transverse distances of the order 
\begin{equation}
b \;\; \sim \;\; 1/M_\pi .
\label{b_chiral}
\end{equation}
Together, Eqs.~(\ref{y_chiral}) and (\ref{b_chiral}) determine the 
parametric region where the pion distribution in the fast--moving 
nucleon is governed by chiral dynamics, and the soft pion can be 
regarded as a ``parton'' in the nucleon's wave function in the
usual sense (see Fig.~\ref{fig:chiral}). 

The condition Eq.~(\ref{y_chiral}) implies that the pion momentum
fraction in the nucleon is parametrically small, $y \ll 1$,
\textit{i.e.}, the soft pion is a ``slow'' parton. 
As a consequence, one can generally neglect the recoil of the spectator 
system and identify the distance $b$ with the separation of the pion from 
the transverse center--of--momentum of the spectator system, 
$r = b/(1 - y)$ \footnote{The relation between the distance of a constituent
from the transverse center--of--momentum, $b$, and its distance from the
center--of--momentum of the spectator system, $r$, can easily be derived for 
the case of a non--interacting system, by starting from the well--known 
expression for the center--of--mass in the rest frame and 
performing a boost to large velocity, taking into account that for the 
non-interacting system the longitudinal momentum fractions are 
given by the ratios of the constituent masses to the total mass of 
the system. A more formal derivation, based on the light--cone components 
of the energy--momentum tensor,
can be found in Ref.~\cite{Burkardt:2002hr}.} \cite{Burkardt:2002hr}. 
This circumstance greatly simplifies the spatial interpretation of 
chiral contributions to the parton densities.
%
% FIGURE
%

Pionic configurations in the nucleon wave function are physically 
meaningful only if the transverse separation of the pion 
and the spectator system is larger than the sum of the intrinsic 
``non--chiral'' sizes of these objects. This basic fact imposes a
limit on the applicability of chiral dynamics, even though the dynamics 
itself may not change dramatically at the limiting distance. 
In order to make the picture of Fig.~\ref{fig:chiral} quantitative,
we have to estimate down to which values of $b$ the concept of
pionic configurations is applicable.

The transverse size of the ``core'' in the nucleon's partonic wave 
function in the valence region ($x \gtrsim 10^{-1}$) can be estimated 
from the transverse axial charge radius of the nucleon, which does not 
receive contributions from the 
pion cloud \cite{Frankfurt:2002ka,Strikman:2003gz}:
\beq
\langle b^2 \rangle_{\rm axial} \;\; = \;\;
{\textstyle\frac{2}{3}} \langle r^2 \rangle_{\rm axial}
\;\; \approx \;\; 0.3 \, \textrm{fm}^2 ,
\label{b2_axial}
\eeq
where the factor $2/3$ results from converting the 3--dimensional
charge radius in the rest frame into the 2--dimensional transverse 
charge radius in the frame where the nucleon is moving fast. 
Identifying the core radius with the transverse RMS radius, 
we obtain
\beq
R_{\rm core} \;\; = \;\; \left[\langle b^2 \rangle_{\rm axial} \right]^{1/2} 
\;\; \approx \;\; 0.55 \, \textrm{fm} .
\label{b_core}
\eeq
Equation~(\ref{b_core}) imposes a numerical lower limit for 
the pion impact parameter, $b$, in pionic configurations.
Note that this number represents a rough estimate, as the interpretation 
of RMS radius in terms of a ``size'' depends on the shape 
of the transverse distribution of partons in the core.
A more refined estimate, which takes into account the intrinsic 
transverse size of the pion as well as the effect of the recoil of the 
spectator system, is obtained by requiring that
$b/(1 - y) > (R_{\rm core}^2 + R_{\pi}^2)^{1/2}$.
Assuming that $R_{\pi}^2$ ranges between zero and $R_{\rm core}^2$,
and anticipating that the typical $y$--values in the pion distribution
at $b \sim R_{\rm core}$ are $y = (1-2) \times M_\pi / M_N \sim 0.2$, 
we obtain $b > 0.44 - 0.62 \, \text{fm}$, in good agreement with 
the estimate of Eq.~(\ref{b_core}).
When considering the nucleon's partonic structure at small $x$
($< 10^{-2}$) the above estimate of the nucleon core size needs to
be modified to account for the non-chiral growth due to diffusion
in the partonic wave function. Also, in this region the transverse
size of the pion itself can grow due to chiral corrections. 
These effects will be discussed separately in Secs.~\ref{subsec:diffusion}
and \ref{subsec:chiral_pion}.

Chiral dynamics produces also configurations in the fast--moving 
nucleon characterized by large longitudinal separations of the 
pion and the spectator system, 
\beq
l \;\; \sim \;\; 1/M_\pi ,
\eeq
with no restriction on $b$. The relevance of these configurations 
for the nucleon's partonic structure cannot be ascertained without
detailed consideration of the effective longitudinal sizes of the 
subsystems and possible coherence effects, and will be discussed
in Sec.~\ref{subsec:longitudinal}. In the following we limit 
ourselves to chiral contributions at large transverse distances.
\subsection{Pion distribution in the nucleon}
\label{subsec:pion_distribution}
\begin{figure}
\includegraphics[width=.4\textwidth]{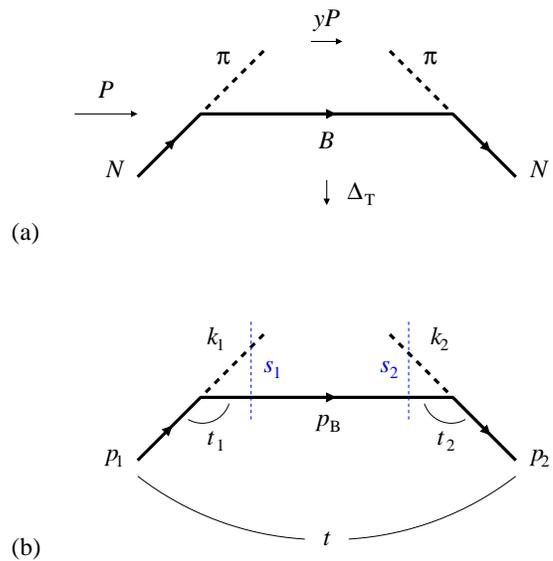}
\caption[]{The pion GPD in the nucleon. (a) Transition matrix element
of the density of pions with longitudinal momentum fraction 
$y \sim M_\pi / M_N$ and transverse momentum 
transfer $|\bm{\Delta}_\perp | \sim M_\pi$, 
Eq.~(\ref{H_pi_number}). (b) Invariants used in modeling finite--size 
effects with form factors. $t_{1, 2}$ are the pion virtualities in the 
invariant formulation, Eq.~(\ref{t_12}); $s_{1, 2}$ the invariant
masses of the $\pi B$ systems in the time--ordered formulation, 
Eq.~(\ref{s_12}).}
\label{fig:gpdpi}
\end{figure}

In its region of applicability defined by Eqs.~(\ref{y_chiral}) 
and (\ref{b_chiral}), the $b$--dependent pion ``parton'' distribution 
can be calculated as the transverse Fourier transform of the 
``pion GPD'' in the nucleon. The latter is defined as the 
transition matrix element of the operator measuring the number density
of pions with longitudinal momentum fraction $y$ in the fast--moving
nucleon, integrated over the pion transverse momenta, and with a transverse 
momentum transfer $\bm{\Delta}_\perp$ to the nucleon 
(see Fig.~\ref{fig:gpdpi}a):
\be
\lefteqn{
\int\!\frac{d^3 k}{(2\pi)^3} \; \delta (y - k_\parallel / P) } &&
\nonumber \\
&\times& \langle \bm{p}_2 | \, 
a_{\pi, a}^\dagger (\bm{k} + \bm{\Delta}/2) \,
a_{\pi, a} (\bm{k} - \bm{\Delta}/2) \, 
| \bm{p}_1 \rangle_{P \rightarrow \infty}
\nonumber \\
&=& (2\pi)^3 \, (2 P) \, \delta^{(3)}(\bm{p}_2 - \bm{p}_1 + \bm{\Delta})
\; H_\pi (y, t) ,
\label{H_pi_number}
\ee
where $p_{1\parallel} = P \rightarrow \infty, \, \Delta_\parallel = 0$,
and 
\beq
t \;\; \equiv -\bm{\Delta}_\perp^2 .
\eeq
Here $a_{\pi, a}^\dagger$ and $a_{\pi, a}$ denote the pion creation and 
annihilation operators, and the sum over isospin projections 
(subscript $a$) is implied. Eq.~(\ref{H_pi_number}) refers to the
helicity--conserving component of the nucleon transition matrix element
($\lambda_2 = \lambda_1$), and $H_\pi (y, t)$ is the corresponding GPD; 
the helicity--flip GPD is defined in analogously but will not be needed 
in the present investigation. In terms of the pion GPD the transverse 
coordinate distribution is then obtained as ($b \equiv |\bm{b}|$)
\be
f_\pi (y, b) &=& \int\frac{d^2 \Delta_\perp}{(2\pi )^2} \; 
e^{-i (\bm{\Delta}_\perp \bm{b})}
\;  H_\pi (y, t) .
\label{f_pi_fourier}
\ee
We note that a manifestly covariant definition of the pion GPD, 
as the matrix element 
of a pionic light--ray operator between nucleon states, was given 
in Ref.~\cite{Strikman:2003gz}; the equivalence of that definition
to Eq.~(\ref{H_pi_number}) is shown by going to the frame where the
nucleon is moving fast and expanding the pion fields in creation and
annihilation operators.
%
% FIGURE
%

The pion GPD in the nucleon implies summation over 
all relevant baryonic intermediate states. Because the pion wavelength 
is assumed to be large compared to the 
typical nucleon/baryon radius, only the lowest--mass excitations can 
effectively contribute to the GPD in the region of Eqs.~(\ref{y_chiral}) 
and (\ref{b_chiral}). We therefore retain only the $N$ and $\Delta$
intermediate states in the sum:
\be
H_\pi &=& H_{\pi N}  + H_{\pi \Delta} , \\
f_\pi &=& f_{\pi N}  + f_{\pi \Delta} . 
\ee
The inclusion of the $\Delta$, whose
mass splitting with the nucleon introduces a non-chiral scale which is
numerically comparable to the pion mass, represents a slight departure 
from strict chiral dynamics but is justified by the numerical importance
of this contribution; \textit{cf.}\ the discussion of the
$N_c \rightarrow \infty$ limit in QCD in Sec.~\ref{sec:largenc}.

To study the properties of the $b$--dependent pion distribution 
at large distances we need a dynamical model which allows us to 
calculate the pion GPD in the relevant region of momenta. Here we 
follow a heuristic approach and start from the simplest possible 
system of pointlike pions and nucleons interacting according to a 
phenomenological Lagrangian. We shall see later how this definition 
can be amended to incorporate finite--size effects. In which region 
the results should be regarded as physical in the light of the 
discussion in Sec~.\ref{subsec:parametric} will be the matter of the 
following investigations.

The pion GPD in the nucleon can be calculated using invariant
perturbation theory, by evaluating the matrix element in 
Eq.~(\ref{H_pi_number}), or, equivalently, the matrix element 
of the pionic light--ray operator of Ref.~\cite{Strikman:2003gz},
using the Feynman rules for pointlike $\pi N$ interactions;
see Ref.~\cite{Strikman:2003gz} for details. 
The resulting Feynman integral is computed
by introducing light--cone coordinates and performing the integral 
over the ``minus'' (energy) component of the loop momentum using 
Cauchy's theorem. Closing the contour 
around the pole of the propagator of the spectator baryon, 
one arrives at a representation in which the
spectator is on mass--shell, and the emitted and absorbed pion
are off mass--shell, with virtualities \footnote{In 
Ref.~\cite{Strikman:2003gz} the pion virtualities were 
denoted by $-s_\pm$. Here we denote them by $t_{1, 2}$, reserving
$s_{1, 2}$ for the invariant masses of the $\pi B$ system, 
Eq.~(\ref{s_12}).}
\beq
t_{1,2} \;\; \equiv \;\; k_{1, 2}^2 
\;\; = \;\; - (\bm{k}_\perp \mp \bar y \bm{\Delta}_\perp / 2)^2 / {\bar y} 
\, + \, t_{{\rm min}} 
\label{t_12}
\eeq
(see Fig.~\ref{fig:gpdpi}b). Here $\bm{k}_\perp$ is the transverse 
momentum of the spectator baryon, 
\beq
\bar y \;\; \equiv \;\; 1 - y ,
\eeq
and 
\beq
t_{{\rm min}} \;\; \equiv \;\;
- \left[ y^2 M_N^2 + y (M_B^2 - M_N^2 ) \right] / {\bar y}
\label{t_min}
\eeq
is the minimum virtuality required by kinematics for a given pion
momentum fraction, $y$. The $\pi N$ and $\pi \Delta$ GPDs are
then obtained as
\be
H_{\pi N} (y, t)  &=& 3 g_{\pi NN}^2 \; I_8 (y, t; M_\pi, M_N) ,
\label{H_pi_N_from_I} \\
H_{\pi \Delta} (y, t) &=& 2 g_{\pi N\Delta}^2 \; 
I_{10} (y, t; M_\pi, M_\Delta) .
\label{H_pi_Delta_from_I}
\ee
Here $g_{\pi NN}$ and $g_{\pi N\Delta}$ are the coupling constants 
in the conventions of Ref.~\cite{Strikman:2003gz} and Appendix~\ref{app:su3},
and the distributions are the isoscalar pion GPDs, corresponding to the 
sum of $\pi^+, \pi^-$ and $\pi^0$ distributions in the proton; 
\textit{cf.}\ Eq.~(\ref{H_pi_number}).
The functions $I_8$ and $I_{10}$ denote the basic transverse 
momentum integrals arising in the calculation of the meson distribution
with intermediate octet and decuplet baryons,
\be
\lefteqn{I_{8, 10} (y, t; M_\pi, M_B)} && \nonumber \\
&\equiv& \frac{y}{4\pi\bar y} \; 
\int\frac{d^2 k_\perp}{(2\pi )^2} \frac{\phi_{8, 10}}
{(t_1 - M_\pi^2) (t_2 - M_\pi^2)} , 
\label{I_8_10}
\ee
where
\be
\phi_8 &\equiv& 
{\displaystyle \frac{1}{2}} \,
\left[ -t_1 - t_2 + \bar y \, t + 2 (M_B - M_N)^2 \right], 
\label{phi_8}
\\[2ex]
\phi_{10} &\equiv& 
{\displaystyle
\frac{1}{24 \, M_N^2 \, M_\Delta^2}} 
\left[ 2 M_\Delta^2 (-t_1 - t_2 + t ) \right. 
\nonumber \\[0ex]
&& \left. + \; (M_N^2 - M_\Delta^2 -t_1) (M_N^2 - M_\Delta^2 -t_2) \right] 
\nonumber \\[0ex]
&& \left. \times \; \left[ 2 (M_\Delta + M_N)^2 - t_1 - t_2 
+ \bar y \, t \right] . \right.
\label{phi_10}
\ee
Note that while the $t_{1, 2}$ of Eq.~(\ref{t_12}) depend on the 
vector $\bm{\Delta}_\perp$, the integral Eq.~(\ref{I_8_10}) 
depends only on $t \equiv -\bm{\Delta}_\perp^2$ because of
rotational invariance in transverse space.

As it stands, the transverse momentum integral in 
Eqs.~(\ref{I_8_10})--(\ref{phi_10})
is divergent. This divergence is related to short--distance contributions 
in the pointlike particle approximation and does not affect the 
chiral long--distance behavior of the $b$--dependent distribution.
Several ways of regularizing this divergence and extracting the 
chiral contribution will be discussed in the following.

The pion GPD can equivalently be evaluated in time--ordered 
perturbation theory, where Fig.~\ref{fig:gpdpi}a is interpreted
as a process where the fast--moving nucleon (momentum $P \gg M_N$) 
makes a transition to a $\pi B$ intermediate state, in which 
we evaluate the operator measuring the density of pions with
longitudinal momentum $yP$, and then back to a nucleon state
whose transverse momentum differs from the original one by 
$\bm{\Delta}_\perp$. In this formulation the 
intermediate particles are on mass--shell, but the energies of the 
$\pi B$ states before and after the operator are different from that of 
the initial/final nucleon state. The invariant masses of the intermediate 
states, which are directly proportional to the energies, are given by
\be
s_{1, 2} &=& (k_{1, 2} + p_B)^2  \\
&=& 
\frac{(\bm{k}_\perp \mp \bm{\Delta}_\perp / 2)^2 + M_\pi^2}{y}
+ \frac{\bm{k}_\perp^2 + M_B^2}{\bar y} 
- \frac{\bm{\Delta}_\perp^2}{4}
\;\;\;\;\;
\label{s_12}
\ee
(see Fig.~\ref{fig:gpdpi}b).
The connection to the invariant formulation is established
by noting that, for given $y$ and $\bm{k}_\perp$,
\beq
\Delta s_{1, 2} \;\; \equiv \;\; s_{1, 2} - M_N^2
\;\; = \;\; \frac{M_\pi^2 - t_{1, 2}}{y} ,
\label{correspondence}
\eeq
whence the denominators in Eq.~(\ref{I_8_10}) 
can also be interpreted as ``energy denominators.'' The minimum
value of the invariant mass difference, $\Delta s_{\rm min}$, 
for given momentum fraction $y$
can be obtained by substituting $t_{1, 2}$ by $t_{\rm min}$,
Eq.~(\ref{t_min}). Both the invariant and the time--ordered 
formulation will be useful for discussing the properties of
the chiral long--distance contribution following from 
Eqs.~(\ref{I_8_10})--(\ref{phi_10}).
\subsection{Large--$b$ asymptotics}
\label{subsec:large_b}
It is instructive to consider the asymptotic behavior of
the distribution of pions for $b \rightarrow \infty$ and fixed $y$. 
It is determined by the leading branch cut singularity of the GPD in 
the $t$--channel and can be calculated by applying the Cutkosky 
rules to the Feynman graphs of Fig.~\ref{fig:gpdpi} with
pointlike vertices \cite{Strikman:2003gz}. The asymptotic 
behavior is of the form
\begin{equation}
f_{\pi B} (y, b) \;\; \propto \;\; 
\frac{e^{\displaystyle -\kappa_{B} b}}{\kappa_{B} b} , 
\label{exponential_largeb}
\end{equation}
where $B = N, \Delta, \ldots$ denotes the intermediate baryon;
the expression applies in principle also to higher--mass 
states, \textit{cf.}\ the discussion below. The decay constant, 
$\kappa_{B}$, depends on the pion momentum fraction, $y$,
and is directly related to the minimum pion virtuality, 
Eq.~(\ref{t_min}), in the invariant formulation,
or the minimum invariant mass difference in the
time--ordered formulation, \textit{cf}.\ Eq.~(\ref{correspondence}):
\beq 
\kappa_B \;\; = \;\; 
2 \left( \frac{M_\pi^2 - t_{\rm min}}{\bar y} \right)^{1/2}.
\label{kappa_B_virtuality}
\eeq
To exhibit the $y$--dependence of the decay constant in the 
parametric region of chiral dynamics, 
$y \sim M_\pi / M_N$, Eq.~(\ref{y_chiral}), 
we set
\beq
y \;\; = \;\; \eta \, M_\pi / M_N , 
\eeq
where the scaling variable, $\eta$, is generally of order unity. 
Substituting Eq.~(\ref{t_min}) into Eq.~(\ref{kappa_B_virtuality})
and dropping terms suppressed by powers of $M_\pi / M_N$, we obtain
\beq
\kappa_{B} \;\; = \;\; 2 \left[ (1 + \eta^2) M_\pi^2
+ \eta \frac{(M_B^2 - M_N^2) M_\pi}{M_N} \right]^{1/2} .
\label{kappa_B_eta}
\eeq
This result has several interesting implications: 
\begin{itemize}
\item[(a)] 
For the nucleon
intermediate state ($B = N$) the second term is zero, and one has
$\kappa_{N} \propto M_\pi$ with a coefficient of order unity and 
depending on $\eta$. In this case the 
$b$--distribution exhibits a ``Yukawa tail'' 
with a $y$--dependent range of the order $1/M_\pi$, as expected. 
\item[(b)] 
For a higher--mass intermediate state ($B \neq N$) the decay
constant is determined by competition of the chiral scale, $M_\pi^2$,
and the non--chiral scale, $(M_B^2 - M_N^2) M_\pi / M_N$. The larger
the $N$--$B$ mass splitting, the smaller $\eta$ has to be for the
chiral scale to dominate. This effect suppresses the contribution 
of higher--mass baryons to $f_{\pi} (y, b)$ at large $b$ and finite 
$\eta$. Note also that the pre-exponential factor, which is 
not shown in Eq.~(\ref{exponential_largeb}) for brevity, 
vanishes $\propto y$ for $y \rightarrow 0$ \cite{Strikman:2003gz}. 
\item[(c)] 
For $\eta \rightarrow 0$ one finds
$\kappa_{B} \rightarrow 2 M_\pi$ irrespective of the $N$--$B$
mass splitting. In this limit the transverse 
``Yukawa tail'' has the range one would naively expect from the
analogy with the 3--dimensional situation. However, this limit
is purely formal, as this region makes a vanishing contribution
to the nucleon's partonic structure at moderate $x$; \textit{cf}.\ 
the discussion in Secs.~\ref{subsec:contribution_nucleon} and
\ref{sec:smallx} below.
\end{itemize}

For pion momentum fractions parametrically of order unity, $y \sim 1$, 
Eq.~(\ref{kappa_B_virtuality}) gives a decay constant 
of the order $\kappa_{B} \sim M_N$. An exponential decay 
with range $\sim 1/M_N$ is not a chiral contribution to the
pion distribution, as is expected, because the values of $y$
lie outside the parametric region of Eq.~(\ref{y_chiral}).
In sum, the large--$b$ asymptotic behavior obtained from the
naive pion distribution with pointlike $\pi N$ couplings 
fully supports the general arguments of Sec.~\ref{subsec:parametric}
concerning the parametric region of the chiral component.

One notes that the characteristic transverse range of the 
chiral contribution of the pion distribution, 
$1/(2 M_\pi) = 0.71 \, \text{fm}$, is numerically
not substantially larger than our estimate of the 
non--chiral ``core'' size, Eq.~(\ref{b_core}).
This shows that an effective field theory approach to chiral dynamics, 
which implicitly assumes that the core has zero size and builds up 
its structure by counter terms, is not practical here, 
and underscores the rationale for our phenomenological approach, 
where finite--size effects are included explicitly.
\subsection{Contribution to nucleon parton densities}
\label{subsec:contribution_nucleon}
The chiral contribution to the nucleon's parton densities is
obtained as the convolution of the pion momentum distribution
in the nucleon with the relevant parton distribution in the pion.
For the gluon, the isoscalar quark/antiquark, and the isovector 
quark/antiquark densities it takes the form \footnote{Equation~(46) 
of Ref.~\cite{Strikman:2003gz} incorrectly writes the convolution formula 
for the antiquark flavor asymmetry with the antiquark density in the 
pion rather than the valence quark density. The correct expression is with 
the valence quark density, \textit{cf.}\ Eq.~(\ref{conv_isovector}). 
This does not affect the conclusions about the large--$N_c$ behavior 
presented in Ref.~\cite{Strikman:2003gz}, which was the sole point of 
the discussion there.} 
\be
\lefteqn{g(x, b)_{\rm chiral}} && \nonumber \\ 
&=& \int_x^1 \frac{dy}{y} \;
\left[ f_{\pi N} + f_{\pi \Delta}\right] (y, b) 
\; g_\pi (z), 
\label{conv_gluon}
\\[3ex]
\lefteqn{
\left[u + d \right] (x, b)_{\rm chiral}
\;\; = \;\; 
\left[ \bar u + \bar d \right] (x, b)_{\rm chiral} 
} && \nonumber \\ 
&=& \int_x^1 \frac{dy}{y} \;
\left[ f_{\pi N} + f_{\pi \Delta}\right] (y, b) 
\; q_\pi^{\text{tot}} (z), 
\label{conv_isoscalar}
\\[3ex]
\lefteqn{
\left[u - d \right] (x, b)_{\rm chiral}
\;\; = \;\; 
\left[ \bar d - \bar u \right] (x, b)_{\rm chiral} 
} && \nonumber \\  
&=& \int_x^1 \frac{dy}{y} \;
\left[ {\textstyle \frac{2}{3} f_{\pi N} - \frac{1}{3} f_{\pi \Delta}} 
\right] (y, b) \; q_\pi^{\text{val}} (z) , 
\label{conv_isovector}
\ee
where
\beq
z \;\; \equiv \;\; x/y 
\label{z_def}
\eeq
is the parton momentum fraction in the pion.
Here $f_{\pi N}$ and $f_{\pi \Delta}$ are the isoscalar pion 
distributions (sum of $\pi^+, \pi^-$ and $\pi^0$) with $N$ and $\Delta$ 
intermediate states in the conventions of 
Refs.~\cite{Koepf:1995yh,Strikman:2003gz} and Appendix~\ref{app:su3};
the isovector nature of the asymmetry, Eq.~(\ref{conv_isovector}),
is encoded in the numerical prefactors. The functions $g_\pi$, 
$q_\pi^{\text{tot}}$, and $q_\pi^{\text{val}}$ are the gluon, 
isoscalar (total), and isovector (valence) quark/antiquark 
densities in the pion,
\be
q_\pi^{\text{tot}} (z) &=& 
\left[ \bar u + \bar d \right]_{\pi\pm, \pi 0} (z)
\;\; = \;\; \left[ u + d \right]_{\pi\pm, \pi 0} (z) 
\nonumber \\
&=& {\textstyle\frac{1}{2}} 
\left[ u + \bar u + d + \bar d \right]_{\pi\pm, \pi 0} (z) ,
\\[2ex]
q_\pi^{\text{val}} (z) &=& 
\pm \left[ \bar d - \bar u \right]_{\pi\pm} (z)
\;\; = \;\; \pm \left[ u - d \right]_{\pi\pm} (z) 
\nonumber \\
&=& \pm {\textstyle\frac{1}{2}} 
\left[ u - \bar u - d + \bar d \right]_{\pi\pm} (z) ;
\label{q_pi_val}
\ee
the latter is normalized as
\beq
\int_0^1 dz \, q_\pi^{\text{val}} (z) \;\; = \;\; 1 .
\label{valence_normalization}
\eeq
The $\pi^0$ does not have a valence distribution because of
charge conjugation invariance, and we assume isospin symmetry.
Note that the parton densities in the pion, as well 
as the result of the convolution integrals in 
Eqs.~(\ref{conv_gluon})--(\ref{conv_isovector}), 
depend on the resolution scale; 
we have suppressed this dependence for brevity. The convolution
formulas for the strange antiquark density and the $SU(3)$--flavor
symmetry breaking asymmetry will be given in Sec.~\ref{sec:decomposition}.

The expressions in Eqs.~(\ref{conv_gluon})--(\ref{conv_isovector}) 
apply to parton momentum fractions of the order $x \sim M_\pi / M_N$
but otherwise not exceptionally small, and transverse distances 
$b \sim 1/M_\pi$. In deriving them we have assumed that the ``decay'' 
of the pion into partons happens locally on the transverse distance 
scale of the chiral $b$--distribution, 
$b \sim 1/M_\pi$ (see Fig.~\ref{fig:chiral}). 
This is justified parametrically, as for the values of $x$ under
consideration the parton momentum fraction in the pion does not 
reach small values ($x < z < 1$ in the convolution integral) 
and one can neglect chiral effects which cause the size of the 
pion itself to grow at small $z$.

To see in which region of $x$ the chiral contribution to the 
isovector antiquark density is localized, it is convenient to write 
the convolution formula Eq.~(\ref{conv_isovector}) in the form
\be
x \left[ \bar d - \bar u \right] (x, b)_{\rm chiral} 
&=& \int_x^1 dy \;
\left[ {\textstyle \frac{2}{3} f_{\pi N} - \frac{1}{3} f_{\pi \Delta}} 
\right] (y, b)
\nonumber \\[1ex]
&\times& \; z q_\pi^{\text{val}} (z) , 
\label{conv_isovector_x} 
\ee
where we have multiplied both sides of Eq.~(\ref{conv_isovector}) 
by $x$ and used Eq.~(\ref{z_def}) on the right--hand side.
Now both functions in the integrand vanish for small arguments: 
$f_{\pi B} (y) \rightarrow 0$ for 
$y \rightarrow 0$, and $z q_\pi^{\text{val}} (z) \rightarrow 0$
for $z \rightarrow 0$. Noting that the valence distribution 
$z q_\pi^{\text{val}} (z)$ is localized around $z \sim 1/2$ at low scales, 
and that the pion momentum distribution is centered around 
$y \sim M_\pi / M_N$, we conclude that the convolution produces a
sea quark distribution in the nucleon centered around values
$x = yz \sim (1/2) \times M_\pi / M_N$, in agreement with the
general expectation. The same argument applies to the bulk of the
chiral isoscalar density, Eq.~(\ref{conv_isoscalar}), which 
arises mainly from the valence quark content of the pion; only at very 
small $x$ the non-valence quarks in the pion produce a distinct contribution.
Note also that the valence quark density in the pion at $z \sim 1/2$
is generated mostly by relatively small--size configurations in the pion,
justifying our approximation of neglecting the intrinsic
transverse size of the pion in the convolution integrals.

One immediately sees from Eqs.~(\ref{conv_isoscalar}) and 
(\ref{conv_isovector}) that the chiral large--distance component
is larger in the isoscalar than in the isovector quark distributions,
because the $N$ and $\Delta$ contributions add in the isoscalar 
sector, Eq.~(\ref{conv_isoscalar}), while they partly cancel in 
the isovector sector, Eq.~(\ref{conv_isovector}) \cite{Koepf:1995yh}. 
This is contrary to the general expectation that chiral effects
manifest themselves mostly in the sea quark flavor asymmetry 
$\bar d - \bar u$. The cancellation between 
$N$ and $\Delta$ contributions in the isovector case becomes
perfect in the large--$N_c$ limit  of QCD and restores
the proper $N_c$ scaling of the isovector distributions; 
see Sec.~\ref{sec:largenc}.

In principle one can use the asymptotic expressions for the
pion distribution in the nucleon, Eqs.~(\ref{exponential_largeb})
and (\ref{kappa_B_virtuality}), to do a numerical estimate of 
the large--distance contribution to the nucleon parton densities
based on Eqs.~(\ref{conv_gluon})--(\ref{conv_isovector}).
This approach was taken in Ref.~\cite{Strikman:2003gz} to 
estimate the chiral contribution to the nucleon's gluonic 
transverse size, $\langle b^2 \rangle_g$, proportional
to the $b^2$--weighted integral of the impact--parameter
dependent gluon density. Because of the weighting with $b^2$
this quantity emphasizes large transverse distances, and the
estimates of the $b$--integrated chiral contribution are relatively 
insensitive to the lower limit in $b$ imposed in the integral
(see also Sec.~\ref{sec:size}).
In the present investigation we are interested in the antiquark 
densities \textit{per se} (not weighted with $b^2$), where there is
no such enhancement of large distances, and estimates of the chiral
contribution are more sensitive to the lower limit in $b$. We 
therefore approach this problem differently, by analyzing the
phenomenological pion cloud model (which incorporates finite--size 
effects) and establishing down to which $b$ the numerical predictions 
are insensitive to the short--distance cutoff (Sec.~\ref{sec:model}). 
The numerical evaluation of the long--distance contribution based 
on Eqs.~(\ref{conv_gluon})--(\ref{conv_isovector}) will then 
be done based on the results of this investigation
(Secs.~\ref{sec:decomposition} and \ref{sec:size}).
\section{Pion cloud model in impact parameter representation}
\label{sec:model}
\subsection{Modeling finite--size effects}
For a quantitative study of the chiral large--distance component 
in the nucleon's partonic structure we need a dynamical model
which allows us to compute the distribution of pions beyond 
its leading asymptotic behavior. In addition, we must address the
question down to which values of $b$ numerical study of this
component is meaningful, in the sense that it is not overwhelmed
by short--distance contributions unrelated to chiral dynamics.
Ultimately, this question can only be answered in a dynamical model 
which smoothly ``interpolates'' between the chiral long--distance 
regime and the effective short--distance dynamics. Here we study 
this question in the framework of the phenomenological pion cloud model, 
where the short--distance dynamics is not treated explicitly, but
modeled by form factors implementing a finite hadronic size 
unrelated to chiral dynamics. This study serves two 
purposes --- it establishes what part of the predictions of the
traditional pion cloud model actually arises from the long--distance 
region governed by chiral dynamics, and it offers a practical way 
of computing this universal long--distance contribution.
 
In the phenomenological pion cloud model, the pion GPD in the
nucleon is defined by the graph of Fig.~\ref{fig:gpdpi},
\textit{cf.}\  Eqs.~(\ref{I_8_10})--(\ref{phi_10}), in which 
now form factors are associated with the $\pi N B$ vertices,
rendering the transverse momentum integral explicitly finite.
Two different schemes to implement these form factors are commonly used 
and have extensively been discussed in the literature. 
One, based on the invariant formulation in which the spectator 
baryon in on mass--shell, restricts the virtualities 
of the exchanged pions by inserting in Eq.~(\ref{I_8_10}) a form factor
\beq
\mathcal{F}\left( \frac{M_\pi^2 - t_{1, 2}}{\Lambda^2_{\text{virt}}} \right)  
\label{ff_virt}
\eeq
for each $\pi N B$ vertex (see Fig.~\ref{fig:gpdpi}b). 
Here $\mathcal{F}(a)$ denotes a function of
finite range which vanishes for $a \rightarrow \infty$; for example,
an exponential, $\exp(-a)$, or the dipole form factor, $(1 + a)^{-2}$. 
These form factors can be compared to those in the well--known meson 
exchange parametrizations of the $NN$ 
interaction, where the exchanged pion is regarded as a virtual 
particle \cite{Machleidt:hj}. The other scheme, based on the time--ordered
formulation, restricts the invariant mass of the $\pi B$ systems 
in the intermediate states by form factors of the type \cite{Zoller:1991cb}
\beq
\mathcal{F}\left( \frac{s_{1, 2} - M_N^2}
{\Lambda^2_{\text{inv.\ mass}}} \right) 
\label{ff_inv}
\eeq
(see Fig.~\ref{fig:gpdpi}b). 
An advantage of this scheme is that it preserves the momentum sum rule 
in the transition $N \rightarrow \pi B$, \textit{i.e.}, the longitudinal 
momentum distribution of the baryon $B$ in the nucleon is given 
by $f_{\pi B} (1 - y)$ for $B = N, \Delta$
\cite{Zoller:1991cb,Melnitchouk:1992yd}. 
The relation between the two different cutoff schemes can easily be
derived from Eq.~(\ref{correspondence}). Effectively,
\beq
\Lambda^2_{\text{virt}} \;\; = \;\; y \, \Lambda^2_{\text{inv.\ mass}} ,
\eeq
\textit{i.e.}, a constant invariant mass cutoff amounts to a 
$y$--dependent virtuality cutoff which tends to zero as $y \rightarrow 0$. 
In the traditional formulation of the pion cloud model, without restriction
to the large--$b$ region, the two schemes lead to rather
different pion momentum distributions. The distributions at large $b$
and $y \sim M_\pi / M_N$, however, are dominated by vanishing pion 
virtualities \textit{viz.}\ invariant mass differences, so that the 
results in the two schemes become effectively equivalent, up to small 
finite renormalization effects. In the following numerical studies
we shall employ the virtuality cutoff as used in Ref.~\cite{Koepf:1995yh}; 
the equivalence of the two schemes for our purposes will be demonstrated 
explicitly in Sec.~\ref{subsec:effective}.

We emphasize that we are interested in the pion cloud model with 
form factors only as a means to identify the chiral large--distance
contribution and delineate the region where it is universal and
independent of the form factors. We do not consider those
aspects of the model related to the fitting of data without
restriction to large distances (tuning of cutoff parameters, 
$\pi N B$ couplings, \textit{etc.}); those have been discussed 
extensively in the literature reviewed in Refs.~\cite{Kumano:1997cy}.
\subsection{Universality at large $b$}
We first consider the dependence of the pion distribution
in the nucleon on the impact parameter, $b$. Specifically, we want
to demonstrate that it reproduces the ``universal'' chiral
behavior Eq.~(\ref{exponential_largeb}) at large $b$, and
investigate for which values of $b$ the distribution is
substantially modified by the form factors. To this end
we calculate the pion GPD by numerical evaluation of
the loop integral, Eq.~(\ref{I_8_10}), with a virtuality
cutoff of the type of Eq.~(\ref{ff_virt}), and perform the transformation 
to the impact parameter representation according to Eq.~(\ref{f_pi_fourier});
useful formulas for the numerical calculation are collected in
Appendix~\ref{app:evaluation}.
Figure~\ref{fig:fb} shows $f_{\pi N}(y, b)$ obtained with
an exponential form factor ($\Lambda_{\text{virt}} = 1.0\, \text{GeV}$, 
a typical value in traditional applications of the 
pion cloud model), as a function of $b$ for $y = 0.07$ and 0.3,
which is 1/2 and 2 times $M_\pi / M_N$, respectively. 
Also shown are the distributions 
obtained with pointlike particles (no form factors), 
in which the loop integral was regularized by subtraction 
at $\bm{\Delta}_\perp^2 = 0$; this subtraction of a 
$\bm{\Delta}_\perp^2$--independent term in the GPD corresponds 
to a modification of the impact parameter distribution
by a delta function term $\propto \delta^{(2)}(\bm{b})$, 
which is ``invisible'' at finite $b$ \cite{Strikman:2003gz}. 
One sees that for $b \gtrsim 0.5 \, \textrm{fm}$ the results of the
two calculations coincide, showing that in this region the pion
distribution is not sensitive to the form factors. Comparison
of different functional forms of the form factor (exponential, dipole) 
also supports this conclusion. Furthermore, we note that for large $b$ 
both distributions in Fig.~\ref{fig:fb} exhibit the universal 
asymptotic behavior derived earlier \cite{Strikman:2003gz}.
%
% FIGURE
%
\begin{figure}
\includegraphics[width=.48\textwidth]{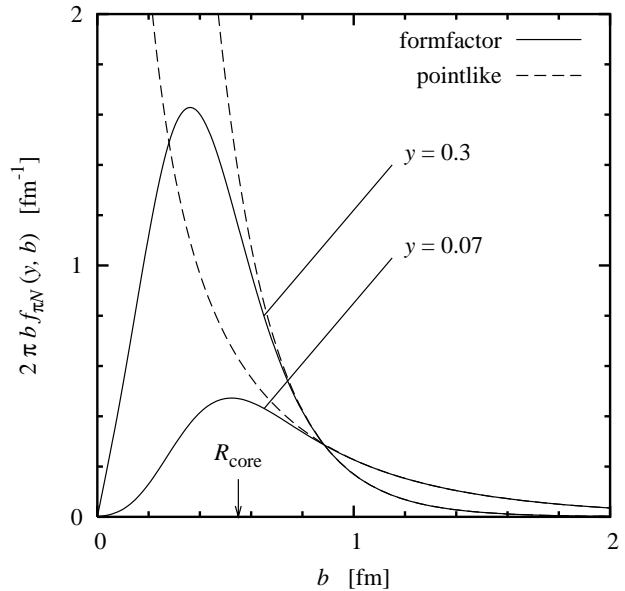} 
\caption[]{The transverse spatial distribution of pions in 
the nucleon, $f_{\pi N} (y, b)$, as a function of $b$, for values
$y = 0.07$ and 0.3. Shown is the radial distribution 
$2\pi b \, f_{\pi N} (y, b)$,
whose integral over $b$ (area under the curve) gives the pion momentum 
distribution. \textit{Solid lines:} Pion cloud model with virtuality
cutoff (exponential form factor, $\Lambda_{\pi N} = 1.0 \, \textrm{GeV}$)
\cite{Koepf:1995yh}. \textit{Dashed line:} Distribution for pointlike
particles, regulated by subtraction at $\bm{\Delta}_\perp^2 = 0$;
the integral over $b$ does not exist in this case. The estimated 
``core'' radius, Eq.~(\ref{b_core}), is marked by an arrow.}
\label{fig:fb}
\end{figure}

It is interesting that the $b$--value where in Fig.~\ref{fig:fb} 
the ``universal'' behavior of $f_{\pi N} (y, b)$ sets in 
is numerically close to the transverse radius of the nucleon's
``core,'' inferred earlier from independent considerations,
$R_{\rm core} \approx 0.55 \, \text{fm}$, \textit{cf.}\ Eq.~(\ref{b_core}).
This shows that the pion cloud model can safely be used to compute 
the large--$b$ parton densities over the entire region defined 
by Eq.~(\ref{b_core}).

%
% FIGURE
%
\begin{figure}
\includegraphics[width=.48\textwidth]{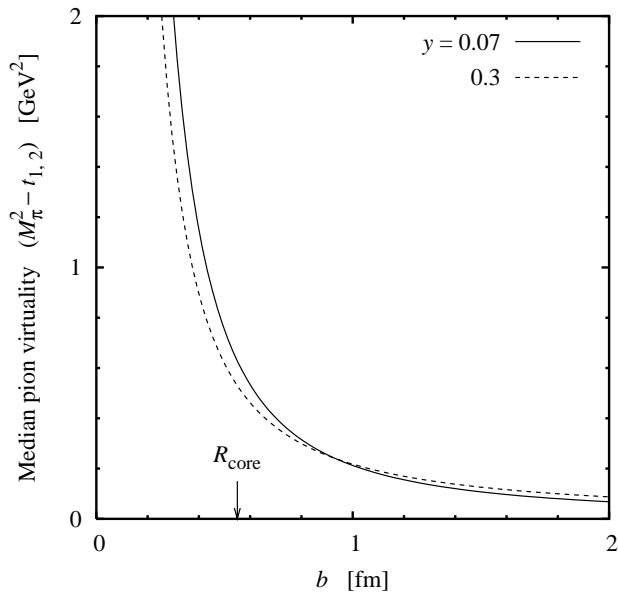}
\caption[]{The median pion virtuality in the unregularized integral,
Eqs.~(\ref{I_8_10})---(\ref{phi_10}), as a function of $b$, 
for $y = 0.07$ (solid line) and $0.3$ (dotted line). 
It is defined as the value of the
virtuality cutoff, $\Lambda^2_{\text{virt}}$, for which 
$f_{\pi N} (y, b)$ reaches half of its value for 
$\Lambda^2_{\text{virt}} \rightarrow \infty$, corresponding
to the unregularized integral.} 
\label{fig:virt}
\end{figure}
Figure~\ref{fig:virt} illustrates the connection between the transverse
distance and the pion virtualities in Eq.~(\ref{I_8_10}) from a 
different perspective. Shown there is the median pion virtuality
in the unregularized loop integral, defined as the value of the 
virtuality cutoff, $\Lambda^2_{\text{virt}}$, for which the
regularized $f_{\pi N} (y, b)$ reaches
half of its value for $\Lambda^2_{\text{virt}} \rightarrow \infty$;
the latter coincides with the value obtained by regularization through
subtraction. The function $f_{\pi N} (y, b)$ is always positive 
when evaluated with an exponential virtuality cutoff, and
monotonously decreasing as a function of $\Lambda^2_{\text{virt}}$, 
so that the median value of $\Lambda^2_{\text{virt}}$ provides a 
sensible measure of the average virtualities in the integral 
Eq.~(\ref{I_8_10}) for given $y$ and $b$. One sees that the
average pion virtualities in the loop strongly decrease with
increasing $b$, indicating the approach to the universal chiral
region. We recall that the leading asymptotic behavior at 
$b \rightarrow \infty$ is determined by quasi--on--shell pions,
\textit{cf.}\ the derivation in Sec.~\ref{subsec:large_b}.
\subsection{Effective pion momentum distribution}
\label{subsec:effective}
%
% FIGURE
%
\begin{figure}
\includegraphics[width=.48\textwidth]{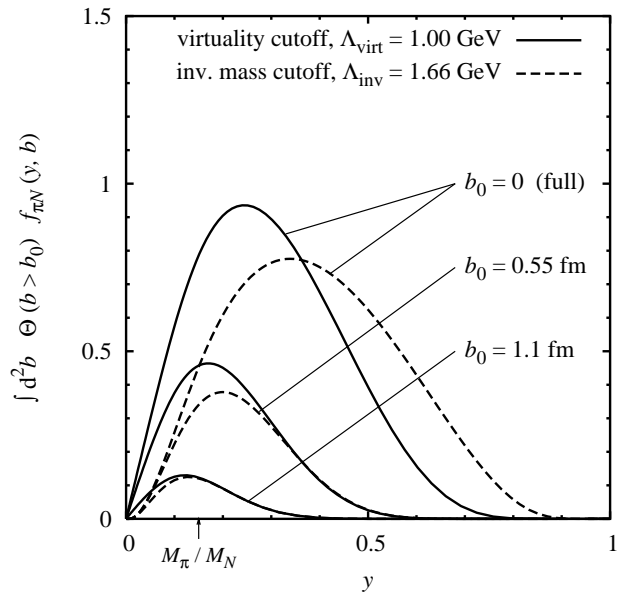}
\caption[]{Effective momentum distribution of pions 
in $\pi N$ configurations with impact parameters $b > b_0$,
Eq.~(\ref{fy_bint}), in the pion cloud model. 
\textit{Solid lines:} Distributions 
obtained with a virtuality cutoff, Eq.~(\ref{ff_virt}) (exponential 
form factor, $\Lambda_{\text{virt}} = 1.0 \, \textrm{GeV}$), for 
$b_0 = 0$ (full integral), $b_0 = 0.55 \, \text{fm}$ and 
$b_0 = 1.1 \, \text{fm}$.
\textit{Dashed lines:} Same for distributions obtained with an
invariant mass cutoff, Eq.~(\ref{ff_virt}) (exponential form factor, 
$\Lambda_{\text{virt}} = 1.66 \, \textrm{GeV}$). The value of
$\Lambda_{\text{virt}}$ was chosen such that it produces the same
total number of pions ($y$--integral) for the full distribution
as the given virtuality cutoff. The value $y = M_\pi / M_N$ 
is indicated by an arrow.}
\label{fig:fy}
\end{figure}
We now want to investigate the distribution of pions at large transverse 
distances as a function of the momentum fraction, $y$. 
In keeping with our general line of approach,
we do this by studying how the momentum distribution of the pion 
cloud model with form factors is modified when a restriction on
the minimum $b$ is imposed. We define the effective momentum
distribution of pions with $b > b_0$ as the integral
\begin{equation}
\int d^2 b \; \Theta (b > b_0) \; f_{\pi B} (y, b) 
\hspace{3em} (B = N, \Delta);
\label{fy_bint}
\end{equation}
for $b_0 = 0$ we recover the momentum distribution of pions
in the traditional usage of the pion cloud model.
Figure~\ref{fig:fy} (solid lines) shows the $b$--integrated
distribution Eq.~(\ref{fy_bint}), obtained with an exponential
virtuality cutoff ($\Lambda_{\text{virt}} = 1.0 \, \text{GeV}$),
for $b_0 = 0$ (full integral) as well as 
$b_0 = 0.55 \, \text{fm}$ and $1.1 \, \text{fm}$, corresponding
to 1 and 2 times the phenomenological core radius, $R_{\text{core}}$.
One sees that the restriction to large $b$--values strongly suppresses 
large pion momentum fractions and shifts the strength of the
distribution toward values of the order $y \sim M_\pi / M_N$, 
in agreement with the general expectations formulated in 
Sec.~\ref{sec:chiral}. From the perspective of the traditional
pion cloud model, the results of Fig.~\ref{fig:fy} 
show that less than half of the pions in that model arise 
from the region $b > R_{\text{core}}$, where the pion cloud can be
regarded as a distinct component of the nucleon wave function

Also shown in Figure~\ref{fig:fy} (dashed lines) are the corresponding
distributions obtained with an invariant mass cutoff, Eq.~(\ref{ff_inv}).
For the sake of comparison the cutoff parameter $\Lambda^2_{\text{inv}}$ 
was chosen here such that it gives the same total number of pions 
($y$--integral) for the ``full'' distributions in which no restriction
on $b$ is imposed ($b_0 = 0$); 
the value of $\Lambda_{\text{inv}} = 1.66 \, \text{GeV}$
thus obtained is within the range considered in phenomenological
applications of the pion cloud model \cite{Melnitchouk:1998rv}.
One sees that the full distributions are quite different for the
virtuality and the invariant mass cutoff, as dictated by the 
relation (\ref{correspondence}). However, when restricted to large
$b$ the $y$--distributions in the two regularization schemes 
become more and more alike, as their strength shifts toward values 
of the order $y \sim M_\pi / M_N$. This explicitly demonstrates the
equivalence of the virtuality and the invariant mass regularization
in the context of our approach, as announced above.
\subsection{Extension to $SU(3)$ flavor}
%
% FIGURE
%
\begin{figure}[b]
\includegraphics[width=.48\textwidth]{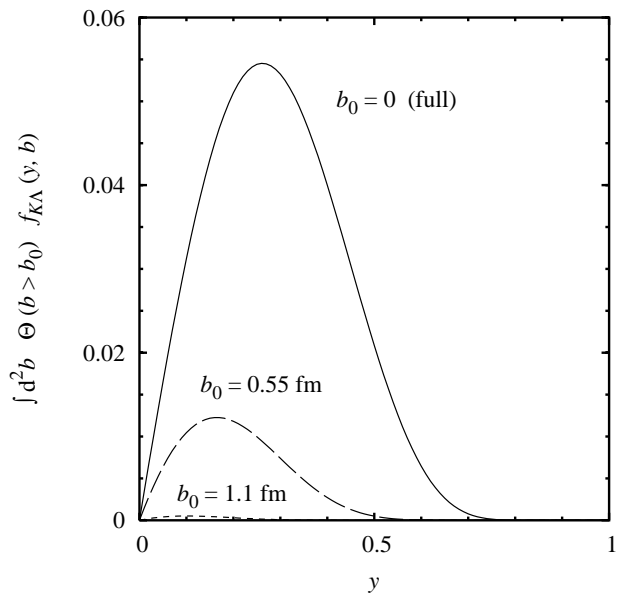}
\caption[]{Effective momentum distribution of kaons in $K\Lambda$ 
configurations with impact parameters $b > b_0$,
\textit{cf.}\ Eq.~(\ref{fy_bint}), in the meson cloud model
with virtuality cutoff (exponential form factor, 
$\Lambda_{\text{virt}} = 1.0 \, \text{GeV}$).
\textit{Solid line:} $b_0 = 0$ (full integral). \textit{Dashed lines:} 
$b_0 = 0.55 \, \text{fm}$ and $b_0 = 1.1 \, \text{fm}$.}
\label{fig:fk_part}
\end{figure}
In our studies of the strange sea and the $SU(3)$--breaking
flavor asymmetry below we shall consider also the contributions 
from $K$ and $\eta$ mesons to the sea quark distributions
at large distances. Because the masses of these mesons are numerically 
comparable to the typical hadronic mass scale (as given, say, by
the vector meson mass), their contributions to the partonic structure 
of the nucleon cannot be associated with chiral dynamics, 
even at large transverse distances. Still, in the context of the
present discussion of the pion cloud model, it is instructive to 
study the distribution of $K$ and $\eta$ in the impact parameter 
representation, and contrast it with that of the $\pi$.

The pseudoscalar octet meson couplings to the nucleon, 
as determined by $SU(3)$ flavor symmetry, and the
definition of their impact parameter--dependent momentum distributions 
are summarized in Appendix~\ref{app:su3}. Significant contributions 
come only from the $K\Lambda$ and $K\Sigma^\ast$ channels.
The large--$b$ behavior of these distributions is formally 
governed by the asymptotic expression, Eqs.~(\ref{exponential_largeb})
and (\ref{kappa_B_virtuality}), with the $\pi$ mass replaced by
the $K$ mass. Figure~\ref{fig:fk_part} shows the numerically
computed effective momentum 
distribution of $K$ in $K\Lambda$ configurations, with and without 
restriction to large $b$, \textit{cf.}\ Eq.~(\ref{fy_bint}), 
which should be compared to the corresponding distributions for the 
$\pi$ in Fig.~\ref{fig:fy}. One sees that the overall magnitude 
of $f_{K\Lambda}$ is substantially smaller than that of $f_{\pi N}$, 
because of the smaller coupling constant (no isospin degeneracy, 
\textit{cf}.\ Appendix~\ref{app:su3}) and the larger meson and intermediate 
baryon mass. More importantly, one notes that the restriction to 
large $b$ suppresses the $K$ distribution much more strongly than 
the $\pi$ distribution; only about $1/5$ of all kaons in the meson
cloud model are located at transverse distances $b > 0.55 \, \text{fm}$, 
and less than $1\%$ are found at $b > 1.1 \, \text{fm}$. 
While hardly surprising, these numbers show clearly that 
the $K$ (and $\eta$) contribution to the partonic structure 
above the nucleon's core radius, $R_{\rm core} = 0.55 \, \text{fm}$ 
is extremely small.
\section{Large--distance component of the nucleon sea}
\label{sec:decomposition}
\subsection{Isovector sea $\bar d - \bar u$}
\label{subsec:isovector}
%
% FIGURE
%
\begin{figure}
\includegraphics[width=.48\textwidth]{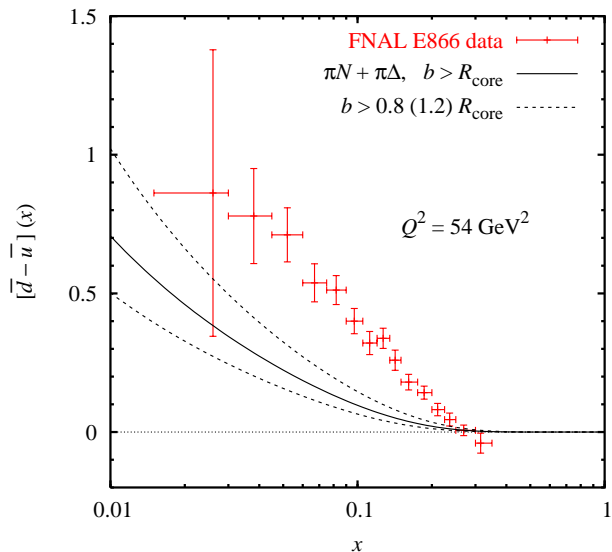} 
\caption[]{\textit{Solid line:} Large--distance contribution to the 
antiquark flavor asymmetry, $[\bar d - \bar u](x)$,
obtained from $\pi N$ and $\pi \Delta$ configurations
restricted to impact parameters $b > R_{\rm core} \
= 0.55 \, \textrm{fm}$, \textit{cf.}\ Eq.~(\ref{fy_bint}). 
\textit{Dotted lines:} Same with $b > 0.8 \; R_{\rm core}$ 
(upper line) and $1.2 \; R_{\rm core}$ (lower line). 
\textit{Data:} Result of analysis of final FNAL E866 
Drell--Yan data \cite{Towell:2001nh}; statistical and systematic 
errors were added in quadrature. All curves and data
points refer to the scale $Q^2 = 54\, \text{GeV}^2$.}
\label{fig:conv}
\end{figure}
We now apply the formalism developed in Secs.~\ref{sec:chiral}
and \ref{sec:model} to study the chiral large--distance 
contributions to the sea quark distributions in the nucleon.
To this end, we evaluate the convolution formulas, 
Eqs.~(\ref{conv_isoscalar})--(\ref{conv_isovector}),
with the $b$--integrated pion distribution, Eq.~(\ref{fy_bint}),
where the lower limit, $b_0$, is taken sufficiently large to 
exclude the model--dependent small--distance region,
\textit{cf.}\ Fig.~\ref{fig:fb}. Our standard value for
this parameter is the phenomenological ``core'' radius, 
Eq.~(\ref{b_core}); variation of this
value  will allow us to estimate the sensitivity of the
results to unknown short--distance dynamics.
While not permitting a complete description of the 
sea quark distributions, our results allow us to quantify
how much comes from the ``universal'' large--distance region,
providing guidance for future comprehensive models of the
partonic structure.

We first consider the isovector antiquark distribution in the proton,
$[\bar d - \bar u](x)$, which experiences only
non--singlet QCD evolution and is largely independent of the
normalization scale. The convolution formula Eq.~(\ref{conv_isovector}) 
involves the valence quark distribution of the pion; the normalization 
of this distribution is fixed by Eq.~(\ref{valence_normalization}), 
and its shape has been determined accurately by fits to the 
$\pi N$ Drell--Yan data; see Ref.~\cite{Gluck:1999xe} and references 
therein. We use the leading--order parametrization of the valence
distribution provided in Ref.~\cite{Gluck:1999xe}; the differences 
to the next--to--leading order parametrization are minor in this case.
Figure~\ref{fig:conv} shows the chiral long--distance contribution 
obtained when $b_0$ is taken to be the phenomenological ``core'' 
radius, $R_{\rm core} = 0.55 \, \textrm{fm}$ (solid line), as well
as the band covered when $b_0$ is changed from this value by
$\pm 20\%$ (dotted lines). Also shown in the figure are the results
of an analysis of the final data from the FNAL E866 
Drell--Yan experiment, presented at a common scale 
$Q^2 = 54 \, \text{GeV}^2$ \cite{Towell:2001nh}. One sees that
the large--distance contribution to the asymmetry is practically zero 
for $x > 0.3$, as expected from the general considerations
of Sec.~\ref{sec:chiral}. At $x\sim 0.1$ the large--distance 
contribution accounts for $\sim 30 \%$ of the measured asymmetry, 
indicating that most of it results from the nucleon's core at small
transverse distances. This conclusion is robust and, as demonstrated
in Sec.~\ref{sec:model}, does not depend on the form factors 
employed in the calculation within the pion cloud model 
(the specific results shown here were obtained with an
exponential virtuality cutoff with 
$\Lambda_{\pi N} = 1.0 \, \textrm{GeV}$ and 
$\Lambda_{\pi\Delta} = 0.8 \, \textrm{GeV}$ \cite{Koepf:1995yh}). 
At small $x$ ($\sim 0.01$) the 
large--distance contribution obtained in our approach comes 
closer to the data; however, the quality of the present data 
is rather poor, and it is difficult to infer the magnitude of 
the required ``core'' contribution by comparing the present estimate 
of the large--distance contribution to the data in this region of $x$.

One sees from Eq.~(\ref{conv_isovector}) that the isovector
antiquark distribution involves strong cancellations between the
contributions from $\pi N$ and $\pi \Delta$ intermediate states.
This is not accidental --- the cancellation between the two
becomes exact in the large--$N_c$ limit of QCD, and is in fact
necessary to restore the proper $N_c$ scaling of the isovector
distribution; see Sec.~\ref{sec:largenc}.
\subsection{Isoscalar sea $\bar u + \bar d$}
The isoscalar light antiquark distribution, $[\bar u + \bar d](x)$,
is subject to singlet QCD evolution and thus exhibits stronger scale 
dependence than the isovector distribution. The convolution formula for
this distribution, Eq.~(\ref{conv_isoscalar}), involves the 
total (singlet) antiquark distribution in the pion, which we 
may write in the form
\beq
q_\pi^{\text{tot}}(z) \;\; = \;\; q_\pi^{\text{val}}(z) 
+ 2 q_\pi^{\text{sea}}(z) ,
\label{q_pi_split}
\eeq
where $q_\pi^{\text{val}}$ is the valence distribution, 
Eq.~(\ref{q_pi_val}), and $q_\pi^{\text{sea}}$ the ``sea''
distribution \footnote{The relation of our conventions for
the pion parton densities to those of Ref.~\cite{Gluck:1999xe} 
(GRS) is $q_\pi^{\text{val}} = \frac{1}{2} v_\pi (\text{GRS}), \;
q_\pi^{\text{sea}} = \bar q_\pi (\text{GRS})$.} 
\beq
q_\pi^{\text{sea}}
\; = \; \bar u_{\pi +}
\; = \; d_{\pi +}
\; = \; u_{\pi -}
\; = \; \bar d_{\pi -} .
\label{q_pi_sea}
\eeq
The pion sea was determined within a radiative parton model analysis, 
supplemented by a constituent quark picture which relates the pion 
to nucleon parton densities, and found to be relatively small
\cite{Gluck:1999xe}. Again, we use the leading--order parametrization
for the parton densities in the pion.

%
% FIGURE
%
\begin{figure}[b]
\includegraphics[width=.48\textwidth]{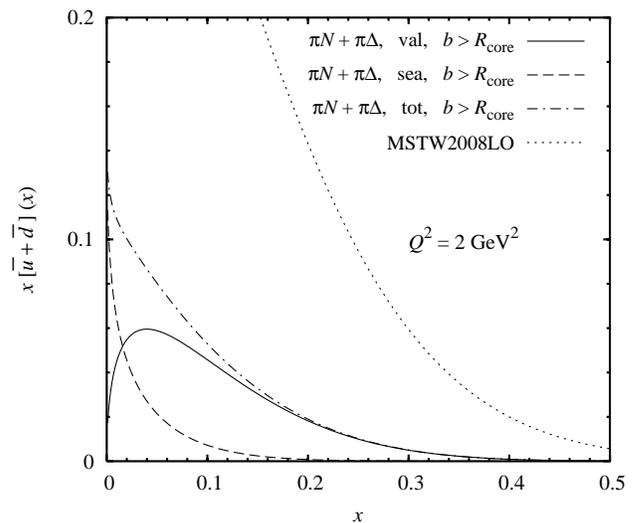} 
\caption[]{\textit{Solid/dashed/dashed--dotted line:} 
Large--distance contribution to the isoscalar antiquark 
density, $x [\bar u + \bar d]$, resulting from $\pi N$ and 
$\pi \Delta$ configurations restricted to 
$b > R_{\rm core} \ = 0.55 \, \textrm{fm}$. The plot shows
separately the contributions arising from the valence, sea, and 
total antiquark density in the pion, \textit{cf.}\ Eq.~(\ref{q_pi_split}).
\textit{Dotted line:} MSTW2008LO leading--order 
parametrization \cite{Martin:2009iq}.}
\label{fig:udbar}
\end{figure}
Figure~\ref{fig:udbar} shows our result for the chiral
large--distance contribution to the isoscalar antiquark distribution,
separately for the valence and sea distributions in the pion as well
as the total, at the scale $Q^2 = 2 \, \text{GeV}^2$. 
One sees that the sea in the pion becomes important only
at $x \ll M_\pi/M_N$, where the antiquark momentum fraction in the
pion can become small, $z \ll 1$. Altogether, the large--distance 
contribution accounts for only $\sim 1/5$ of the total 
$\bar u + \bar d$ in the nucleon at $x \sim 0.1$.

The antiquark distribution obtained from $\pi B \, (B = N, \Delta)$
configurations cannot be larger than the total antiquark distribution 
in the nucleon, which includes the radiatively generated sea. 
The large--distance contribution calculated in our approach easily 
satisfies this theoretical constraint, as can be seen from the
comparison with the parametrization obtained in the recent 
leading--order global fit of Ref.~\cite{Martin:2009iq} (MSTW20008LO),
see Fig.~\ref{fig:udbar}.
We note that the traditional pion cloud model without restriction 
to large $b$, which generates pions with transverse momenta of the 
order $\sim 1 \, \text{GeV}$ and virtualities $\sim 1 \, \text{GeV}^2$, 
produces an isoscalar sea which comes close to saturating the empirical 
$\bar u + \bar d$ at large $x$ with the usual range
of parameters parameters, and can even overshoot it for 
certain choices \cite{Koepf:1995yh,Melnitchouk:1998rv}.
The restriction of $\pi B$ configurations to large $b$ in our approach 
solves this problem in a most natural way.
\subsection{Strange sea $s, \bar s$} 
\label{subsec:sbar}
The strange sea ($s, \bar s$) in the nucleon at large distances
has two distinct components. One is the chiral component,
arising from $s$ and $\bar s$ in the pion in $\pi N$ and
$\pi \Delta$ configurations. It is given by a similar convolution
formula as the isoscalar sea, $\bar u + \bar d$, 
Eq.~(\ref{conv_isoscalar}),
\beq
\bar s (x, b)_{\rm chiral} 
\;\; = \;\; \int_x^1 \frac{dy}{y} \;
\left[ f_{\pi N} + f_{\pi \Delta}\right](y, b) \; 
\bar s_\pi (z) ,
\eeq
and similarly for $s$, where $\bar s_\pi (z)$ and $s_\pi (z)$
are the strange (anti--) quark distributions in the pion.
Assuming that the sea in the pion is mostly generated 
radiatively \cite{Gluck:1999xe},
we take them to be equal and proportional to the non--strange
sea in the pion, Eq.~(\ref{q_pi_sea}),
\beq
\bar s_\pi (z) 
\; = \; s_\pi (z) \; = \; q_\pi^{\text{sea}}(z) .
\label{pion_sea_su3}
\eeq
The other component comes from valence $\bar s$ quarks in 
$KY (Y = \Lambda, \Sigma, \Sigma^\ast)$ and $\eta N$ configurations 
in the nucleon. Because the masses of these mesons are numerically 
comparable to the typical hadronic mass scale (as given, say, by
the vector meson mass), their contribution to the partonic structure 
of the nucleon cannot strictly be associated with chiral dynamics, 
even at large transverse distances. We include them in our numerical 
studies because (a) it is instructive to contrast their contribution
to those of $\pi N$ and $\pi \Delta$; (b) they contribute to $\bar s$
only and could in principle generate different $x$--distributions
for $s$ and $\bar s$, as suggested by the model of 
Ref.~\cite{Brodsky:1996hc} (we shall comment on this model below).
The couplings of the octet mesons to the nucleon, as determined
by $SU(3)$ symmetry and the quark model value of the $F/D$ ratio,
as well as the definitions of the corresponding 
meson momentum distributions are summarized in Appendix~\ref{app:su3}. 
The contribution of $K$ and $\eta$ to $\bar s(x, b)$ in the proton 
is obtained as
\be
\bar s (x, b) &=& \int_x^1 \frac{dy}{y} \; \left\{ {\textstyle\frac{2}{3}}
f_{\eta N} (y, b) \; \bar s_\eta (z) \right.
\nonumber \\
&+& \left. \left[ f_{K\Lambda} + f_{K\Sigma} + f_{K\Sigma^\ast} 
\right] (y, b) \; \bar s_K (z) \right\} ,
\label{conv_sbar}
\ee
where the factor $2/3$ accounts for the probability of the $\eta$
to be in a configuration with a valence $\bar s$ quark
(we assume a pure octet state of the $\eta$ and do not take into 
account singlet--octet mixing, as the $\eta$ contribution turns out 
to be negligibly small anyway). The functions $\bar s_\eta (z)$ and 
$\bar s_K (z)$ are the normalized momentum distributions of 
$\bar s$ in $\eta$ and $K$,
\beq
\int_0^1 dz \, \bar s_{\eta, K} (z) \;\; = \;\; 1 .
\eeq
Assuming $SU(3)$ symmetry, we will approximate these distributions
by the valence quark distribution in the pion,
\beq
\bar s_{\eta, K} (z) \;\; 
\approx \;\; q_\pi^{\text{val}} (z) .
\eeq
We again use the leading--order parametrization of Ref.~\cite{Gluck:1999xe} 
for the valence quark density in the pion.
Numerical evaluation of the meson distributions shows that 
the contributions from $\eta N$ and $K\Sigma$ in Eq.~(\ref{conv_sbar})
are negligible because of their relatively small coupling; 
we retain only the $K\Lambda$ and $K\Sigma^\ast$ terms in 
the following. 

%
% FIGURE
%
\begin{figure}
\includegraphics[width=.48\textwidth]{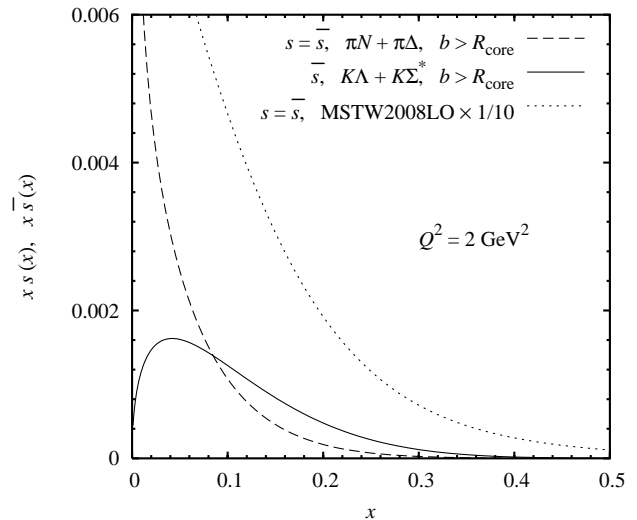}
\caption[]{\textit{Dashed line:} Large--distance contribution to 
the strange sea in the nucleon, $s = \bar s$, from $\pi N$ and 
$\pi \Delta$ configurations ($b > R_{\rm core} \ = 0.55 \, \textrm{fm}$).
\textit{Solid line:} Large--distance contribution to $\bar s$ 
from $K\Lambda$ and $K\Sigma^\ast$ configurations, involving the 
valence strange quark distribution in the kaon;
$K\Sigma$ and $\eta N$ are numerically negligible.
\textit{Dotted line:} MSTW2008LO leading--order parametrization 
of the total $s = \bar s$ \cite{Martin:2009iq}, multiplied by $1/10$ 
for easier comparison.}
\label{fig:sbar}
\end{figure}
Figure~\ref{fig:sbar} shows the different large--distance contributions to 
the strange sea, integrated over $b > R_{\text{core}} = 0.55 \, \text{fm}$.
One sees that for $x > 0.1$ the large--distance strange sea is mostly 
$\bar s$ coming from the valence $\bar s$ in $K\Lambda$ and $K\Sigma^\ast$
configurations; the precise magnitude of this contribution 
is sensitive to the lower limit in $b$, \textit{cf.}\ 
Fig.~\ref{fig:fk_part}. For $x < 0.1$ the large--distance strange sea 
in the nucleon originates mostly from the strange sea in the pion
in $\pi B \, (B = N, \Delta)$ configurations, which contributes
equally to $s$ and $\bar s$. The different mechanisms result in
$s(x) \neq \bar s(x)$ for the large--distance component of the
strange sea. However, the overall magnitude of the large--distance
component represents only $\sim 1/20$ of the empirically determined 
average strange sea, $\frac{1}{2} \left[ s + \bar s \right](x)$ 
\cite{Martin:2009iq}, so that one cannot draw any conclusions about 
the $x$--distributions of the total $s$ and $s$ in the nucleon 
from the large--distance component. Note that the large--distance 
component at $b > R_{\text{core}}$ represents a much smaller fraction 
of the total sea in the case of $s$ and $\bar s$ than for 
$\bar u + \bar d$, at least in the region $x \gtrsim 0.01$.

There are significant differences between the leading--order
parametrizations of $[s + \bar s](x)$ obtained in the global fits of 
Refs.~\cite{Martin:2009iq} and \cite{Gluck:2007ck}; up to a factor 
$\sim 2$ at $x = 0.1$. However, this does not change our basic 
conclusion, that the large--distance $s$ and $\bar s$ are only a 
small fraction of the total. Also, some of the next--to--leading
order fits by several groups \cite{Martin:2009iq,Lai:2007dq}
have begun to extract information on 
the shapes of $s(x)$ and $\bar s (x)$ individually, by incorporating 
neutrino scattering data which discriminate between the 
two. The difference $[s - \bar s](x)$ 
is very poorly determined by the existing data, and the fits serve 
mostly to limit the range of allowed values. We note that 
our approach to large--distance contributions and the convolution 
formulas of Eqs.~(\ref{conv_gluon})--(\ref{conv_isovector})
remain valid also for next--to--leading order parton densities,
if the parton densities in the pion are taken to be the 
next--to--leading order ones. In the present study we restrict 
ourselves to the leading order, because at this order the parton
densities are renormalization--scheme--independent and possess
a simple probabilistic interpretation, and because the present 
comparison of our results with the data does not warrant high accuracy.

We would like to comment on the approach of Ref.~\cite{Brodsky:1996hc},
where the shapes of $s(x)$ and $\bar s(x)$ were investigated in a 
light--front wave function model with $K\Lambda$ components, whose 
amplitude was adjusted to fit the observed total strange sea, 
$[s + \bar s](x)$. As just explained, our results show that only a 
very small fraction of the total $s$ and $\bar s$ sea arise from 
transverse distances $b > R_{\rm core} \approx 0.55 \, \text{fm}$ 
where the notion of meson--baryon components in the nucleon wave function 
is physically sensible. Even in the traditional meson cloud model without 
restriction to large $b$, $K\Lambda$ configurations with standard 
form factors \cite{Koepf:1995yh,Holtmann:1996be} 
would account only for $\sim 1/4$ of the present value of 
$s + \bar s$ \cite{Martin:2009iq}. This shows that the assumption 
of saturation of the strange sea by $K\Lambda$ configurations made 
in Ref.~\cite{Brodsky:1996hc} would require a $KN\Lambda$ coupling
$\sim 2$ times larger than the $SU(3)$ value and is not realistic. 
While we see indications for $s(x) \neq \bar s(x)$ in the large--distance 
contribution, and certainly nothing requires the shapes to be equal,
the magnitude of the effect cannot be reliably predicted on
the basis of the model of  Ref.~\cite{Brodsky:1996hc}.
\subsection{Flavor asymmetry $\bar u + \bar d - 2\bar s$}
%
% FIGURE
%
\begin{figure}[b]
\includegraphics[width=.48\textwidth]{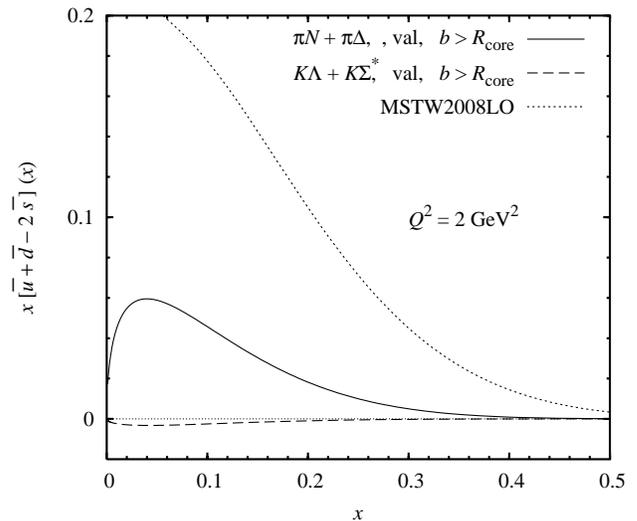} 
\caption[]{\textit{Solid line:} Large--distance contribution
to the antiquark $SU(3)$ flavor asymmetry asymmetry 
in the nucleon $\bar u + \bar d - 2 \bar s$ from valence
$\bar u + \bar d$ in the pion in $\pi N$ and $\pi \Delta$ configurations,
\textit{cf.}\ Fig.~\ref{fig:udbar} 
($b > R_{\rm core} \ = 0.55 \, \textrm{fm}$).
\textit{Dashed line:} Contribution from the valence $\bar s$ in
the kaon in $K\Lambda$ and $K\Sigma^\ast$ configurations, 
\textit{cf.}\ Fig.~\ref{fig:sbar}.
\textit{Dotted line:} Leading--order parametrization
of Ref.~\cite{Martin:2009iq}.}
\label{fig:qbar8}
\end{figure}
\label{subsec:qbar3}
The antiquark $SU(3)$ flavor asymmetry 
$\bar u + \bar d - 2\bar s$ is a non--singlet
combination of the isoscalar non--strange and strange sea, which
exhibits only weak scale dependence. Since we assume $SU(3)$
flavor symmetry of the sea quarks in the pion, Eq.~(\ref{pion_sea_su3}),
only the valence $\pi$ and $K$ components of $\bar u + \bar d$
and $\bar s$ enter in this combination (we neglect the $\eta N$
and $K\Sigma$ contributions):
\be
\left[ \bar u + \bar d - 2 \bar s \right] (x, b)
&\approx& \int_x^1 \frac{dy}{y} \left[ f_{\pi N} + f_{\pi \Delta}
- 2 f_{K\Lambda}
\right. \nonumber \\
&& \left. - 2 f_{K\Sigma^\ast} \right](y, b) 
\;\; \bar q_\pi^{\text{val}} (z) .
\label{conv_qbar8}
\ee
The large--distance contribution to this asymmetry is shown in
Fig.~\ref{fig:qbar8}. One sees that the asymmetry overwhelmingly
results from the valence $\bar u$ and $\bar d$ content of the
pion in $\pi N$ and $\pi\Delta$ configuration; the $\bar s$
in the kaon of $K\Lambda$ and $K\Sigma^\ast$ contributes only
at the level of $< 10\%$ of the pion. Overall, the large--distance
contribution accounts for $\sim 1/3$ of the observed $SU(3)$ flavor
asymmetry at $x \sim 0.1$.
\section{Transverse size of nucleon}
\label{sec:size}
\subsection{Transverse size and GPDs}
An interesting characteristic of the nucleon's partonic 
structure is the average squared transverse radius of the 
partons with given longitudinal momentum fraction $x$.
It is defined as
\beq
\langle b^2 \rangle_f (x) \;\; \equiv \;\; 
\frac{\displaystyle\int d^2 b \; b^2 \; f(x, b)}{\displaystyle f(x)}
\hspace{3em} (f = q, \bar q, g),
\label{b2_def}
\eeq
where $f(x, b)$ is the impact parameter--dependent distribution
of partons, related to the total parton density by
\beq
\displaystyle\int d^2 b \; f(x, b) \;\; = \;\; \displaystyle f(x) .
\eeq
The average is meaningful thanks to the positivity of 
$f(x, b)$ \cite{Burkardt:2002hr,Pobylitsa:2002iu}. 
Physically, Eq.~(\ref{b2_def}) 
measures the average transverse size of configurations in the nucleon 
wave function contributing to the parton density at given $x$.
The transverse size implicitly depends also on the scale, $Q^2$;
this dependence arises from the DGLAP evolution of the impact--parameter
dependent parton distribution and was studied 
in Ref.~\cite{Frankfurt:2003td}.

The average transverse quark/antiquark/gluon size of the nucleon is 
directly related to the $t$--slope of the corresponding 
nucleon GPD at $t = 0$,
\beq
\langle b^2 \rangle_f (x) \;\; = \;\; 
4 \, 
\frac{\partial}{\partial t}
\left[ \frac{H_f (x, t)}{H_f (x, 0)} \right]_{t = 0} .
\eeq
Here $H_f (x, t) \equiv H_f (x, \xi = 0, t)$ denotes the ``diagonal'' 
GPD (zero skewness, $\xi = 0$), with $H_f (x, 0) = f(x)$,
which is related to the impact parameter--dependent 
distribution as ($b \equiv |\bm{b}|$)
\beq
H_f (x, t = -\bm{\Delta}_\perp^2) \;\; = \;\; 
\int d^2 b \; e^{-i (\bm{\Delta}_\perp \bm{b})} \;
f (x, b) .
\label{H_f_fourier}
\eeq
For a general review of GPDs and their properties we refer to
Refs.\cite{Goeke:2001tz}.
\subsection{Transverse size from hard exclusive processes}
\label{subsec:hard_exclusive}
By virtue of its connection with the GPDs, the transverse size 
of the nucleon is in principle accessible experimentally, 
through the $t$--slope of hard exclusive processes, 
\beq
\gamma^\ast(Q^2) + N \rightarrow M + N \hspace{2em} 
(M = \text{meson}, \gamma, \ldots),
\nonumber
\eeq
at $Q^2 \gg 1\, \text{GeV}^2$ and $|t| \lesssim 1\, \text{GeV}^2$, 
whose amplitude can be calculated using QCD factorization 
and is proportional to the nucleon GPDs. In general, 
such processes require a longitudinal momentum transfer to the nucleon 
and probe the ``non--diagonal'' GPDs ($\xi \neq 0$), so that
the connection between the observable $t$--slope and the transverse size
can be established only with the help of a GPD parametrization
which relates the distributions at $\xi \neq 0$ to those at $\xi = 0$.
The connection becomes simple in the limit of high--energy 
scattering, $\xi \approx x_B / 2 \ll 1$, where the GPDs probed
in the hard exclusive process can be related to 
the diagonal ones in a well--controlled approximation; see 
Refs.~\cite{Frankfurt:1997ha,Shuvaev:1999ce} for details. 
In this approximation the amplitudes for light vector meson
electroproduction at small $x$ ($\phi, \rho$) and
heavy vector meson photo/electroproduction ($J/\psi, \Upsilon )$ 
are  proportional to the diagonal gluon GPD, and thus
\beq
(d\sigma/dt)^{\gamma^\ast N \rightarrow V + N}
\;\; \propto \;\; H_g^2 (x = x_B, t) .
\label{dsigma_gluon}
\eeq
The gluonic average transverse size can be directly inferred 
from the relative $t$--dependence of the differential cross section
\beq
\langle b^2 \rangle_g
\;\; = \;\; 4 \, \frac{\partial}{\partial t} \left[
\frac{d\sigma/dt \, (t)}{d\sigma/dt \, (0)} \right]^{1/2}_{t = 0} .
\eeq
The universal $t$--dependence of exclusive $\rho^0$ and $\phi$ 
electroproduction at sufficiently large $Q^2$ and exclusive 
$J/\psi$ photo/electroproduction, implied by Eq.~(\ref{dsigma_gluon}),
is indeed observed experimentally and represents an important
test of the approach to the hard reaction mechanism; 
see Ref.~\cite{Levy:2009ek} for a recent compilation of results. 

Experimental information on the nucleon's gluonic size 
and its dependence on $x$ comes mainly from the extensive data 
on the $t$--dependence of exclusive
$J/\psi$ photo/electroproduction, measured in the HERA 
H1 \cite{Aktas:2005xu} and ZEUS \cite{Chekanov:2004mw} experiments, 
as well as the FNAL E401/E458 \cite{Binkley:1981kv} and other 
fixed--target experiments; see Ref.~\cite{Jung:2009eq}
for a recent summary. The $t$--dependence of the cross section 
measured in the HERA experiments is well described by an 
exponential,
\beq
(d\sigma/dt)^{\gamma N \rightarrow J/\psi + N}
\;\; \propto \;\; \exp (B_{J/\psi} t) ,
\eeq
and assuming that this form is valid near $t = 0$, the nucleon's
average gluonic transverse size is obtained as
\beq
\langle b^2 \rangle_g \;\; = \;\; 2 B_{J/\psi} .
\label{b2_gluon_slope}
\eeq
For a more accurate estimate, the measured $t$--slope is reduced
by $\sim 0.3 \, \text{GeV}^{-2}$ to account for the finite
size of the produced $J/\psi$. The exponential slope measured by 
H1 at $\langle W\rangle = 90 \, \text{GeV}$
is $B_{J/\psi} = 4.630 \pm 0.060 {}^{+0.043}_{-0.163} \, \text{GeV}^{-2}$
\cite{Aktas:2005xu}, and ZEUS quotes a value of 
$B_{J/\psi} = 4.15 \pm 0.05 {}^{+0.30}_{-0.18} \, \text{GeV}^{-2}$
\cite{Chekanov:2004mw}. The central values correspond to a transverse 
gluonic size at $x \sim 10^{-3}$ in the range 
$\langle b^2 \rangle_g = 0.31-0.35 \, 
\text{fm}^2$, substantially smaller than the transverse size of the 
nucleon in soft hadronic interactions.
It is also found that the gluonic size increases with $\log (1/x)$ 
with a coefficient much smaller than the soft--interaction Regge slope,
\textit{cf.}\ the discussion in Sec.~\ref{subsec:diffusion}.

Comparatively little is known about the quark size of the nucleon 
at small $x$. As explained above, light vector meson production 
at small $x$ couples mainly to the gluon GPD. 
Interesting new information comes 
from the $t$--dependence of deeply--virtual Compton scattering (DVCS) 
recently measured at HERA. The H1 experiment \cite{Aaron:2007cz} 
obtained an exponential 
slope of $B_\gamma = 5.45 \pm 0.19 \pm 0.34 \, \text{GeV}^{-2}$ 
by measuring $t$ through the photon transverse momentum; 
larger by one unit than the $J/\psi$ slope measured by the 
same experiment. ZEUS \cite{Chekanov:2008vy} extracted a DVCS slope
of $B_\gamma = 4.5 \pm 1.3 \pm 0.4 \, \text{GeV}^2$ by measuring
the transverse momentum of the recoiling proton, again larger than
the $J/\psi$ slope measured by that experiment; however, the 
exponential fit to the ZEUS data is rather poor and the extracted 
$B_\gamma$ has large errors. We note that in both experiments the 
$B_\gamma$ values were 
determined by an exponential fit over the entire measured region of
$t$ and thus reflect the average $t$--dependence, not directly the
slope at $t = 0$. Still, the data provide some indication that the 
$t$--slope of DVCS at $t = 0$ is larger than that of $J/\psi$
production (the $Q^2$ in the DVCS experiments here 
are comparable to the effective scale in $J/\psi$ photoproduction, 
$Q^2_{\rm eff} \approx 3 \, \text{GeV}^2$). In leading--order 
(LO) QCD factorization, the DVCS amplitude is proportional to
the singlet quark GPDs, and the $t$--slope of this process
is directly related to the nucleon's singlet quark size, 
\beq
\langle b^2 \rangle_{q + \bar q} \;\; = \;\; 2 B_\gamma ,
\label{b2_dvcs_slope}
\eeq 
\textit{cf.}\ Eq.~(\ref{b2_gluon_slope}). One would thus conclude that
\beq
\langle b^2 \rangle_{q + \bar q} 
\;\; > \;\; \langle b^2 \rangle_g .
\label{b2_quark_vs_gluon}
\eeq
At next--to--leading order (NLO) the DVCS amplitude also involves 
the gluon GPD, and substantial cancellation is found between the gluon 
and singlet quark contributions to the amplitude. This 
cancellation amplifies the effect of a difference in 
$\langle b^2 \rangle_{q + \bar q}$ and $\langle b^2 \rangle_g$ 
on the DVCS $t$--slope. Because the gluon contribution is negative 
and cancels $\sim 1/2$ of the quark contribution \cite{Freund:2001hm}, 
the relative change in the slope should be $\sim 2$ times larger than 
the relative change in the average transverse sizes which caused 
it \cite{Lim:2006xu}.

The quark transverse size of the nucleon in the valence quark region
is measured in hard exclusive processes at Jefferson Lab, in particular
with the 12 GeV Upgrade. In this kinematics the skewness of the GPDs
needs to be taken into account ($\xi \neq 0$), and the analysis relies
on GPD parametrizations. It is interesting that the $t$--slope
of $\rho^0$ production measured in the recent CLAS 
experiment \cite{Morrow:2008ek} seems to be compatible with the 
Regge--based GPD parametrization of Ref.~\cite{Guidal:2004nd}
(however, it is presently unclear how to describe the absolute 
cross section within this framework). A detailed phenomenological 
study of the transverse distribution of valence quarks, based on
parton densities and form factor data, was performed in
Ref.~\cite{Diehl:2004cx}.
\subsection{Chiral contribution}
We now want to study the contribution of the chiral large--distance 
region, $b \sim 1/M_\pi$, to the nucleon's average transverse size.
Adopting a two--component description, we define
\be
\langle b^2 \rangle_f \!
&=& \! \frac{\displaystyle \int \!\! d^2b \, b^2 \; 
\left[ f(x, b)_{\rm core} + \Theta (b > b_0) \,
f(x, b)_{\rm chiral} \right]}{\displaystyle f(x)}
\nonumber \\
&\equiv& \langle b^2 \rangle_{f, \, {\rm core}} 
\; + \; \langle b^2 \rangle_{f, \, {\rm chiral}} .
\label{core_cloud}
\ee 
Here $f(x, b)_{\rm core}$ denotes the parton density arising
from average configurations in the nucleon, 
distributed over transverse distances $b \sim R_{\rm core}$.
The function $f(x, b)_{\rm chiral}$ is the chiral component
of the parton distribution, extending over distances $b \sim 1/M_\pi$.
Following the same approach as above, we integrate it over $b$
with a lower cutoff, $b_0$, of the order of the core radius,
Eq.~(\ref{b_core}); the sensitivity of the results to the precise
value of $b_0$ will be investigated below. Note that in 
Eq.~(\ref{core_cloud}) the $b^2$--weighted integral in the numerator 
is computed in two separate pieces, while the denominator in both 
cases is the total parton density (core plus chiral) at the given 
value of $x$; the $\langle b^2 \rangle_{f, \, {\rm chiral}}$
thus defined represents the contribution of the chiral component to
the overall transverse size of the nucleon, not the ``intrinsic''
size of the chiral component alone.

The ``core'' contribution to $\langle b^2 \rangle$ was estimated
in Sec.~\ref{subsec:parametric} and Ref.~\cite{Strikman:2003gz},
by relating it to the slope of the nucleon's axial form 
factor, which does not receive contributions from the pion cloud:
\beq
\langle b^2 \rangle_{\rm core} 
\;\; \approx \;\; {\textstyle\frac{2}{3}} \, \langle r^2 \rangle_{\rm axial}
\;\; = \;\; 0.3 \, \text{fm}^2 .
\label{bulk_axial}
\eeq
We have already used this result to fix the short--distance cutoff
in the integral over the chiral contribution. 
A more quantitative determination of the ``non--chiral'' transverse 
sizes of the nucleon, including the differences between quarks, 
antiquarks and gluons and their $x$--dependence, requires a dynamical
model of the nucleon which smoothly interpolates between small and large 
distances and will be the subject of a separate study. 
Here we focus on the chiral contribution,
$\langle b^2 \rangle_{f, \, {\rm chiral}}$, which can be calculated
in a model--independent manner; we compare it to the  
``generic'' core size given by Eq.~(\ref{bulk_axial}), keeping in
mind that the latter may have a richer structure than reflected
by this simple estimate.

The chiral contribution to the transverse size, Eq.~(\ref{core_cloud}),
is obtained by calculating the $b^2$--weighted integral of the 
$b$--dependent pion momentum distribution in the nucleon 
studied in Sec.~\ref{sec:model}, \textit{cf.}\ Eq.~(\ref{fy_bint}), 
and substituting the 
result in the convolution formula for the nucleon parton density,
Eq.~(\ref{conv_isoscalar}) \textit{et seq.} Useful formulas for the 
numerical evaluation of the $b^2$--weighted integrals are
presented in Appendix~\ref{app:evaluation}. Because of the 
weighting factor $b^2$, the chiral contribution to the transverse 
size is much less sensitive to unknown short--distance dynamics
(\textit{i.e.}, to the cutoff $b_0$) than the contribution to the 
parton density itself, and thus represents a much more interesting
quantity for studying effects of chiral dynamics in the partonic
structure. Furthermore, the $b^2$--weighted integral can reliably be 
computed using the asymptotic form of the distribution of pions at 
large $b$, Eq.~(\ref{exponential_largeb}), as was done in 
Ref.~\cite{Strikman:2003gz}. We can use this to estimate analytically
the sensitivity of $\langle b^2 \rangle_{f, {\rm chiral}}$ to the 
lower limit, $b_0$. Evaluating the integral 
\be
I &\equiv& 
\int d^2 b \; \Theta (b > b_0) \; b^2 \; f_{\pi N} (y, b) 
\ee
with the asymptotic expression Eq.~(\ref{exponential_largeb}), and 
taking the logarithmic derivative with respect to $b_0$, we obtain
\be
- \frac{b_0}{I} \; \frac{\partial I}{\partial b_0} 
%&=& \left. \frac{\lambda^3}{2 + 2 \lambda + \lambda^2}
%\; \right|_{\textstyle\lambda = \kappa_N b_0} \\[1ex]
&\approx& \frac{1}{5} \; \ll \; 1
\hspace{2em} (y = M_\pi / M_N),
\ee
where we have used Eq.~(\ref{kappa_B_eta}) for $\kappa_N$ and 
$b_0 = R_{\rm core} = 0.55\, \text{fm}$. 
This shows that the sensitivity of $\langle b^2 \rangle_{\rm chiral}$
is indeed low --- a 20\% change in $b_0$ causes only a 4\% change in
$\langle b^2 \rangle_{f, {\rm chiral}}$.

%
% FIGURE
%
\begin{figure}
\includegraphics[width=.48\textwidth]{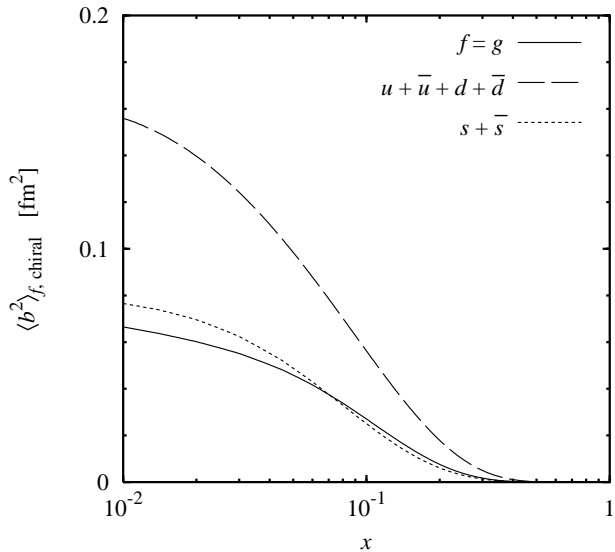} 
\caption[]{The chiral large--distance contribution to the 
nucleon's average transverse size, $\langle b^2 \rangle_f$, as
defined by Eq.~(\ref{core_cloud}), as a function of $x$
($Q^2 = 3 \, \text{GeV}^2$). \textit{Solid line:} Gluonic size ($f = g$),
\textit{cf.}\ Ref.~\cite{Strikman:2003gz}. 
\textit{Dashed line:} Singlet quark size ($f = u + \bar u + d + \bar d$)
\textit{Dotted line:} Strange quark size ($f = s + \bar s$).
In all cases, the curves show the sum of contributions from $\pi N$ and 
$\pi \Delta$ configurations. 
\label{fig:b2}}
\end{figure}
Our results for the chiral contribution to the nucleon's average
transverse size and its dependence on $x$ are summarized 
in Fig.~\ref{fig:b2}, for the scale $Q^2 = 3 \, \text{GeV}^2$.
The curves shown are the sum of contributions from $\pi N$ and 
$\pi \Delta$ configurations; heavier mesons make negligible
contributions at large distances. For reasons of consistency
the nucleon parton densities in the denominator of 
Eq.~(\ref{core_cloud}) were evaluated using the older
parametrization of Ref.~\cite{Gluck:1998xa}, which served as
input to the analysis of the pion parton distributions of
Ref.~\cite{Gluck:1999xe}. The following features are
worth mentioning:
\begin{itemize}
\item[(a)] 
The chiral contribution to the transverse size
is practically zero above $x \sim M_\pi / M_N \sim 0.1$
and grows rapidly as $x$ drops below this value,
in agreement with the basic picture described in Sec.~\ref{sec:chiral}.
The rise of $\langle b^2 \rangle_{f, {\rm chiral}}$ with decreasing $x$ 
is more pronounced than that of the parton density itself because 
the former quantity emphasizes the contributions from large distances.
\item[(b)] The singlet $u$-- and $d$--quark size grows more 
rapidly with decreasing $x$ than the gluonic radius. This has a 
simple explanation: the quark/antiquark density in the pion sits 
at relatively large momentum fractions $z \sim 0.5$, while the 
gluon density in the pion requires $z < 0.1$ to be sizable;
because $z = x/y$ in the convolution integral, and the
pion momentum fractions are of the order $y \sim M_\pi / M_N$, 
the relevant values of $z$ are reached much earlier 
for the quark than for the gluon as $x$ decreases below $M_\pi / M_N$.
Thus, the chiral large--distance contribution suggests 
that the transverse quark size of the nucleon at $x \lesssim 0.01$
is larger than the transverse gluon size, \textit{cf.}\ 
Eq.~(\ref{b2_quark_vs_gluon}). The difference between the 
chiral contribution to the average sizes at $x = 0.01$ is
\beq
\langle b^2 \rangle_{q + \bar q, {\rm chiral}} 
- \langle b^2 \rangle_{g, {\rm chiral}} 
\;\; = \;\; 0.09 \, \text{fm}^2 .
\eeq
Assuming identical core sizes for the quark and gluon distribution,
this would correspond to a difference of the leading--order
DVCS and $J/\psi$ $t$--slopes, 
\textit{cf.}\ Eqs.~(\ref{b2_gluon_slope}) and (\ref{b2_dvcs_slope}),
\beq
B_\gamma - B_{J/\psi} \;\; = \;\; 1.1 \, \text{GeV}^2 ,
\label{b_diff}
\eeq
well consistent with the HERA results summarized in 
Sec.~\ref{subsec:hard_exclusive}. It should be remembered that
the chiral prediction, Eq.~(\ref{b_diff}), is for the exact $t$--slope 
of the cross section at $t = 0$, while the HERA results represent 
effective slopes, obtained by fitting the empirical $t$--dependence 
over the measured range; the comparison may be affected
by possible deviations of the true $t$--dependence from the
exponential shape. More quantitative conclusions would require 
detailed modeling of the core contributions to the transverse size, 
which themselves can grow with decreasing $x$ due to diffusion,
see Sec.~\ref{subsec:diffusion}.
\item[(c)] The chiral contribution to the transverse strange quark
size of the nucleon closely follows that to the gluonic size. 
This is natural, as $s + \bar s$ is mostly generated radiatively,
by conversion of gluons into $s\bar s$, in both the pion and 
the nucleon.
\end{itemize}
\section{Pion cloud and large--$N_c$ QCD}
\label{sec:largenc}
The relation of the chiral component of the large--$b$ parton 
densities to the large--$N_c$ limit of QCD 
is a problem of both principal and practical significance.
First, in the large--$N_c$ limit QCD is expected to become equivalent
to an effective theory of mesons, in which baryons appear as solitonic
excitations, establishing a connection with the phenomenological notion
of meson exchange. Second, our calculations show that contributions
from $\Delta$ intermediate states are numerically large, and the
large--$N_c$ limit provides a conceptual framework which allows one
to treat $N$ and $\Delta$ states on the same footing and relate their 
masses and coupling constants. We now want to verify that the 
large--distance component of the nucleon's partonic structure, 
calculated from phenomenological pion exchange, exhibits
the correct $N_c$--scaling required of parton densities in QCD
(\textit{cf.}\ also the discussion in Ref.~\cite{Strikman:2003gz}).

The general $N_c$ scaling of the unpolarized quark densities 
in the nucleon in QCD is of the form \cite{Diakonov:1996sr}
\be
g(x) &\sim & N_c^2 \times \text{function} (N_c x) ,
\label{nc_gluon} \\[0ex]
[u + d](x), \; 
[\bar u + \bar d](x) &\sim & N_c^2 \times \text{function} (N_c x) ,
\label{nc_isoscalar}
\\[0ex]
[u - d](x), \;
[\bar u - \bar d](x) &\sim & N_c \times \text{function} (N_c x) ,
\label{nc_isovector}
\ee
where the scaling functions are stable in the large--$N_c$ limit
but can be different between the various distributions.
Equations~(\ref{nc_isoscalar}) and (\ref{nc_isovector}) 
were derived by assuming non--exceptional configurations ($x \sim N_c^{-1}$)
and fixing the normalization of the scaling function from the 
lowest moments of the parton densities, \textit{i.e.}, from the
conditions that the total number of quarks scale as $N_c$, and the
nucleon isospin as $N_c^0$. The transverse coordinate--dependent parton
distributions should generally scale in the same manner as the total 
densities, Eqs.~(\ref{nc_isoscalar}) and (\ref{nc_isovector}), 
as the nucleon radius is stable in the large--$N_c$ limit
(this applies even to the nucleon's chiral radii, because
$M_\pi \sim N_c^0$).

Turning now to the pion cloud contribution to the parton 
densities at large $b$, it follows from the expressions of
Eqs.~(\ref{H_pi_N_from_I})--(\ref{phi_10}) and their Fourier
transform, Eq.~(\ref{f_pi_fourier}), that the $b$--dependent
distributions of pions in the nucleon scale as \cite{Strikman:2003gz}
\beq
f_{\pi N}(y, b), f_{\pi\Delta} (y, b) 
\;\; \sim \;\; N_c^2 \times \text{function} (N_c x) .
\label{f_pi_scaling}
\eeq
This behavior applies to values $y \sim M_\pi / M_N \sim N_c^{-1}$
and values $b \sim N_c^0$, corresponding to $|t| \sim N_c^0$ in
the pion GPD. In arriving at Eq.~(\ref{f_pi_scaling}) we have used
that $M_N, M_\Delta \sim N_c$; that $g_{\pi N N} \sim N_c^{3/2}$, 
as implied the Goldberger--Treiman relation; and that 
$g_{\pi N \Delta}$ scales in the same way as $g_{\pi NN}$.
Equation~(\ref{f_pi_scaling}) states that the momentum distribution 
of pions in the nucleon at large $N_c$ scales like that of isoscalar 
quarks or gluons. At the same time, the parton densities in the 
pion scale as
\beq
g_\pi (z), \, q_\pi (z) \;\; \sim \;\; \text{function} (z) ,
\eeq
where $z \sim N_c^0$ in typical configurations; that is,
they have no explicit $N_c$ dependence at large $N_c$. 
One thus concludes that the $N_c$--scaling 
of the convolution integral for the pion cloud contribution
to the nucleon's antiquark densities, for both $B = N$ and $\Delta$ 
intermediate states, is
\beq
\int_x^1 \frac{dy}{y} f_{\pi B}(y, b) \; q_\pi (z) 
\;\; \sim \;\; N_c^2 \times \text{function} (N_c x) .
\eeq
This correctly reproduces the general $N_c$--scaling of the 
isoscalar quark and gluon distribution, Eq.~(\ref{nc_isoscalar}),
where the $N$ and $\Delta$ contributions are added, 
\textit{cf.}\ Eq.~(\ref{conv_isovector}). However, it may seem
that the pion cloud contribution at large $b$ cannot reproduce
the $N_c$ scaling of the isovector distribution, Eq.~(\ref{nc_isoscalar}),
which is suppressed by one power. The paradox is resolved when 
one notes that in the large--$N_c$ limit the $N$ and $\Delta$
become degenerate, 
\beq
M_\Delta - N_N \;\; \sim \;\; N_c^{-1} ,
\eeq
and their couplings are related by \cite{Adkins:1983ya}
\beq
g_{\pi N \Delta} \;\; = \;\; {\textstyle\frac{3}{2}} \, g_{\pi N N} .
\label{g_largenc}
\eeq
Using these relations one has
\beq
f_{\pi \Delta} (y, b) \;\; = \;\; 2 \, f_{\pi N} (y, b)
\hspace{3em} (y \sim N_c^{-1}),
\label{nc_f}
\eeq
as can be seen from Eqs.~(\ref{H_pi_N_from_I})--(\ref{phi_10})
and Eq.~(\ref{f_pi_fourier}), keeping in mind that
$t, t_1, t_2 \sim N_c^0$ in the region of interest.
By virtue of Eq.~(\ref{nc_f}) the $N$ and $\Delta$ contributions
at large $N_c$ cancel exactly in the isovector convolution integral, 
Eq.~(\ref{nc_isovector}), ensuring that the result has the proper
$N_c$--scaling behavior as Eq.~(\ref{nc_isovector}).

In sum, our arguments show that the pion exchange contribution
at large $b$ is a legitimate part of the nucleon's 
partonic structure in large--$N_c$ QCD, exhibiting the same scaling
behavior as the corresponding ``average'' distributions.
The inclusion of $\pi \Delta$ configurations at the same
level as $\pi N$ is essential because they reproduce the proper 
$N_c$--scaling of the isovector distributions, and because
they make numerically sizable contributions --- 
twice larger than $\pi N$ --- to the isoscalar distributions.

In Ref.~\cite{Strikman:2003gz} we have shown that the isoscalar
large--$b$ pion distribution in the nucleon 
$\left[ f_{\pi N} + f_{\pi \Delta} \right](y, b)$
obtained from phenomenological soft--pion exchange, 
can equivalently be computed in the chiral soliton picture of 
the nucleon at large $N_c$, as a certain longitudinal Fourier 
transform of the universal classical pion field of the soliton at 
large transverse distances. Extending this connection
to the isovector pion distribution, $\left[ \frac{2}{3} f_{\pi N} 
- \frac{1}{3} f_{\pi \Delta} \right] (y, b)$, 
which is suppressed in the large--$N_c$ 
limit, remains an interesting problem for further study. In particular,
this requires establishing the connection between soft--pion
exchange and the collective rotations of the classical soliton.
\section{Small $x$--regime and longitudinal distances}
\label{sec:smallx}
\subsection{Growth of core size through diffusion}
\label{subsec:diffusion}
In our studies so far we have focused on chiral contributions
to the nucleon's partonic structure at moderately small momentum
fractions, $x \gtrsim 10^{-2}$, 
which arise from individual $\pi B \, (B = N, \Delta)$
configurations in the nucleon wave function. When considering smaller
values of $x$ several effects must be taken into account which 
potentially limit the validity of the present approximations.

One of them is the growth of the transverse size of ``average'' 
partonic configurations in the nucleon due to diffusion. 
Generally, the partons at small $x$ are decay products of partons 
at larger $x$; the decay process has the character of a random walk in 
transverse space and leads to a logarithmic growth of the transverse 
area occupied by the partons:
\be
\langle b^2 \rangle_{\rm parton} &=&
\langle b^2 \rangle_{\rm parton} (x_0) \; + \; 
4 \, \alpha'_{\rm parton} \, 
\ln (x_0 / x) 
\nonumber \\
&& 
(x < x_0 \sim 10^{-2}).
\ee
The rate of growth --- the effective Regge slope, 
$\alpha'_{\rm parton}$ --- depends on the type of parton
and generally decreases with increasing scale 
$Q^2$, because
higher $Q^2$ increase the effective transverse momenta in the
decay process \cite{Frankfurt:2003td}. Measurements of the
energy dependence of the $t$--slope of exclusive $J/\psi$ production 
at HERA H1 and ZEUS \cite{Aktas:2005xu,Chekanov:2004mw} indicate that 
the rate of growth for gluons at a scale $Q^2 \approx 3 \, \text{GeV}^2$
is approximately $\alpha'_g \approx 0.14 \, \text{GeV}^{-2}$
\footnote{The value quoted here corresponds to the arithmetic mean 
of the parametrizations of $\alpha'_{J/\psi}$ quoted by the HERA 
H1 \cite{Aktas:2005xu} and ZEUS \cite{Chekanov:2004mw}
experiments; see Ref.~\cite{Jung:2009eq} for details.} 
significantly smaller than the rate of growth of the transverse
nucleon size in soft hadronic interactions, 
$\alpha'_{\rm soft} \approx 0.25 \, \text{GeV}^{-2}$. 
Using the former value as a general measure of the rate of growth 
of the nucleon's transverse size due to diffusion, we estimate 
that at $Q^2 \approx 3 \, \text{GeV}^2$ the transverse size of 
the ``core'' increases from $R_{\rm core}^2 = 
0.3\, \text{fm}^2$ at $x = 10^{-2}$ 
to $0.35 \, (0.4)\, \text{fm}^2$ at $x = 10^{-3} \, (10^{-4})$. 
In principle this effect pushes the region of $\pi B$ configurations 
governed by chiral dynamics out to larger $b$ as $x$ decreases. 
However, the rate of growth at this scale is still rather small, 
leaving ample room for such configurations in the region
$x > 10^{-3}$. Note that at lower scales the rate of growth is
larger; studies based on DGLAP evolution show that $\alpha'_g$ 
approaches the soft value at 
$Q^2 \sim 0.4\, \text{GeV}^2$ \cite{Frankfurt:2003td}.
\subsection{Chiral corrections to pion structure}
\label{subsec:chiral_pion} 
Another effect which needs to be taken into account at small $x$
are modifications of the parton density in the pion itself due 
to chiral dynamics. The same mechanism as discussed above for the
nucleon in principle works also in the pion itself --- the pion
can fluctuate into configurations containing a ``slow'' pion and
a two--pion spectator system. When evaluated in chiral perturbation
theory, the momentum fraction of the slow pion relative to its parent
in such configurations
is of the order $y(\text{$\pi$ in $\pi$}) \sim M_\pi / (4\pi F_\pi)$, 
where $F_\pi$ is the pion decay constant, and $4\pi F_\pi$ represents
the generic short--distance scale appearing in the context of the
renormalization of the loop integrals. Such contributions to the
parton density and the GPD in the pion were recently computed
in an all--order resummation of the leading logarithmic approximation
to chiral perturbation theory \cite{Kivel:2008ry}, which does not 
require knowledge of the higher--order terms in the chiral Lagrangian.
For the nucleon parton densities this mechanism could become important 
for $x \ll 10^{-2}$, where the effective parton momentum fractions 
in the pion can reach small values $z \lesssim 0.1$. In the present
study we restrict ourselves to nucleon parton densities 
at $x \gtrsim 10^{-2}$, for which the convolution integrals are
dominated by ``non--chiral'' values of $z$.
The incorporation of such corrections to the partonic structure 
of the pion and extension of the present nucleon structure 
calculation toward smaller $x$ remains an interesting problem 
for future study. In particular, it should be investigated 
how the expressions derived in the leading--log approximation of
chiral perturbation theory compare to a ``single--step'' calculation 
of pion structure including finite mass and size (form factors),
along the lines done here for the nucleon.
\subsection{Chiral dynamics at large longitudinal distances}
\label{subsec:longitudinal} 
In our studies in Secs.~\ref{sec:chiral}--\ref{sec:size} we considered 
chiral contributions to the nucleon's partonic structure at large 
transverse distances, which arise from $\pi B$ configurations at 
large transverse separations, $b \sim 1/M_\pi$.
As already indicated in Sec.~\ref{subsec:parametric}, there is 
in principle another class of $\pi B$ configurations governed by chiral 
dynamics, namely those corresponding to large longitudinal separations
in the nucleon rest frame,
\beq
l \;\; \sim \;\; 1/M_\pi ,
\label{r_longitudinal}
\eeq
and arbitrary values of the transverse separation, down to $b = 0$. 
We now want to discuss in which region of $x$ such configurations 
can produce distinct contributions to the partonic structure. 

The main limitation in admitting $\pi B$ configurations of the type 
Eq.~(\ref{r_longitudinal}) as part of the partonic structure arises
from the possible longitudinal overlap of the relevant partonic 
configurations in the pion and the ``core.'' To determine the region 
where this effect plays a role, it is useful to consider instead of the 
parton densities the structure function for $\gamma^\ast N$ scattering 
and appeal to the notion of the coherence length of the virtual photon. 
Contributions to the partonic structure of the type of the convolution 
integrals of Eqs.~(\ref{conv_gluon})--(\ref{conv_isovector}) 
correspond to the impulse approximation of $\gamma^\ast N$ scattering, 
which requires that the coherence length of the
process be smaller than the longitudinal distance between the
constituents, so that interference effects can be neglected;
see \textit{e.g.}\ Ref.~\cite{Frankfurt:1981mk}. Generally, 
the coherence length for $\gamma^\ast N$ scattering in the nucleon 
rest frame is given by
\beq
l_{\rm coh} \;\; = \;\; (2 M_N x)^{-1} ,
\eeq
where $M_N$ is the nucleon mass and $x \approx Q^2/W^2 \ll 1$ the 
Bjorken variable; $W$ is the center--of--mass energy of the
scattering process. Thus, one would naively 
think that in scattering from a $\pi N$ system with longitudinal
separation $\sim (2 M_\pi)^{-1}$ coherence effects set in if 
$x < M_\pi / M_N \sim 0.1$. However, this argument neglects the
fact that in the fast--moving nucleon the pion carries only a fraction
of the order $y \sim M_\pi / M_N \sim 0.1$ of the nucleon's momentum,
so that the effective center--of--mass energy for $\gamma^\ast\pi$ 
scattering is actually lower by this factor, and the coherence length 
smaller by this factor, than in the $\gamma^\ast N$ process.
Interference effectively takes place only when the coherence length
for both scattering on the pion and on the baryon in the
$\pi B$ configuration is $\sim (2 M_\pi)^{-1}$, which requires
\beq
x \;\; \lesssim \;\; 0.01 . 
\eeq 
For larger values of $x$ coherence effects are small, and there is 
in principle room for a chiral component of the partonic structure
at small $b$ and longitudinal distances $\sim 1/M_\pi$. In order
to calculate this component one would need to model the finite--size
effects limiting the longitudinal extension of the pion and the 
spectator system, which is related to the ``small--$x$ behavior'' 
of the parton densities of the respective systems. We leave this 
problem to a future study. Interestingly, this could result in 
partial ``readmission'' of the small--$b$ component of the pion 
cloud model which was excluded in the present study, potentially
affecting \textit{e.g.}\  the comparison with the measured flavor 
asymmetry $\bar d - \bar u$ in Sec.~\ref{subsec:isovector}.

We note that the interference effects in scattering from $\pi B$ 
configurations described here are large in the region in which
chiral corrections to the structure of the pion would become
important, \textit{cf.}\ the discussion in Sec.~\ref{subsec:chiral_pion}.
An interesting question is whether in the chiral perturbation theory
approach these effects come into play already at the level of the leading 
logarithmic approximation \cite{Kivel:2008ry}, or only at the level 
of subleading or finite terms.
\section{Summary and outlook}
\label{sec:summary}
The transverse coordinate representation based on GPDs represents
a most useful framework for studying the role of chiral dynamics 
in the nucleon's partonic structure. It allows one to identify the 
parametric region of the chiral component 
($x \lesssim M_\pi / M_N, b \sim 1/M_\pi$) and provides
a practical scheme for calculating it in a model--independent way. 
Let us briefly summarize the main results of our investigation.
\begin{itemize}
\item[(a)] 
The contributions from $\pi B \, (B = N, \Delta)$ 
configurations to the parton distributions become independent
of the $\pi NB$ form factors at transverse distances 
$b \gtrsim 0.5 \, \textrm{fm}$, and thus can be associated
with universal chiral dynamics. The lower limit in $b$ 
approximately coincides with the nucleon's core radius, 
$R_{\rm core} = 0.55 \, \textrm{fm}$, inferred previously from
other phenomenological considerations.
\item[(b)] 
Only $\sim 1/3$ of the measured antiquark flavor asymmetry
$[\bar d - \bar u](x)$ at $x > 10^{-2}$ comes from the 
large--distance region $b > R_{\rm core}$, showing that 
most of it resides in the nucleon's core at small transverse distances.
The traditional pion cloud model, which attempts to explain the
entire asymmetry from pionic contributions, gets most of the
effect from small $b$ where the concept of $\pi B$ configurations
is not applicable.
\item[(c)] 
The isoscalar antiquark distribution $[\bar u + \bar d](x)$ 
obtained from pions at large $b$ remains safely below the 
total antiquark distribution determined by QCD fits to 
deep--inelastic scattering data, leaving room for the
(non--perturbatively and perturbatively generated) 
antiquarks in the core. This naturally solves a problem of 
the traditional pion cloud model, where the pionic contribution
can saturate or even exceed the total antiquark density
for certain non-exceptional parameter values.
\item[(d)] 
The strange sea quark distributions, $s(x)$ and $\bar s(x)$,
overwhelmingly sit at small transverse distances, $b < R_{\rm core}$.
Neither chiral ($\pi N, \pi\Delta$) nor $K\Lambda$ configurations
at large $b$ account for more than a few percent of the empirical
$s + \bar s$. The predictions of Ref.~\cite{Brodsky:1996hc} for the 
$x$--dependence of $s(x)$ and $\bar s(x)$ from $K\Lambda$ 
fluctuations rely on the region where the concept of 
distinct meson--baryon configurations is not applicable and 
require a probability of $K\Lambda$ fluctuations several times 
larger than what is obtained from the standard $SU(3)$ couplings.
\item[(e)] 
The pionic contributions to the nucleon's transverse size, 
$\langle b^2 \rangle$, are much less sensitive to short--distance 
dynamics than those to the parton distributions themselves,
and thus furnish a new set of clean chiral observables.
The large--distance contributions to the nucleon's singlet quark 
size at $x < 0.1$ are larger than those to the gluonic size,
suggesting that 
$\langle b^2 \rangle_{q + \bar q} > \langle b^2 \rangle_g$,
in agreement with the pattern of $t$--slopes of deeply--virtual 
Compton scattering and exclusive $J/\psi$ production measured 
at HERA and FNAL. 
\end{itemize}

In the present study we have limited ourselves to the universal 
large--distance contributions to the partonic structure,
which are governed by soft pion exchange and can be calculated
in a model--independent way. A complete description should include
also a model of the short--distance part, which actually carries
most of the parton densities. One way of combining the two would
be a two--component picture, in which the constituents in the ``core''
act as a source of the chiral pion fields which propagate out to
distances $\sim 1/M_\pi$. Such an approach would be very effective
if the characteristic transverse sizes of the ``cloud'' and the 
``core'' were numerically very different. However, this is not the
case --- the characteristic range of two--pion exchange $1/(2 M_\pi) 
= 0.71 \, \text{fm}$ is numerically not much larger than our
estimate of the ``core'' size, $R_{\rm core} = 0.55 \, \text{fm}$.
Another approach, which appears more promising, is based on the idea of 
a smooth ``interpolation'' between the chiral large--distance dynamics 
and the short--distance regime. In particular, the effective
theory of Ref.~\cite{Diakonov:1984tw}, which is based on the 
large--$N_c$ limit of QCD, uses constituent quarks as interpolating 
degrees of freedom; it is valid in a wide region, from distances of the 
order $\sim 1/M_\pi$ down to distances
of the order $\rho \approx 0.3 \, \text{fm}$ --- the range of the 
non-perturbative chiral--symmetry breaking forces in the QCD vacuum.
It leads to a picture of the nucleon as quarks bound by a 
self--consistent pion field (chiral quark--soliton model)
\cite{Diakonov:1987ty}, which is fully field--theoretical and 
relativistic and provides a very good description of the nucleon's 
quark and antiquark densities, including subtle effects such as the 
sea quark flavor asymmetry and its polarization \cite{Diakonov:1996sr}.
The results of Ref.~\cite{Strikman:2003gz} and the present work
(see in particular Sec.~\ref{sec:largenc}) show that this large--$N_c$
description of nucleon structure is equivalent to phenomenological 
soft--pion exchange at large transverse distances, thanks to the
universality of chiral dynamics; it thus, in a sense, contains the 
result of the present work as a limiting case. Using this large--$N_c$ 
picture as a script to model the impact parameter--dependent parton
densities at all $b$, would certainly be an interesting problem for
further study.

Direct experimental study of the chiral component of the nucleon's 
partonic structure through hard exclusive processes at $x < 0.1$ 
would be possible with a future electron--ion collider (EIC). 
The simplest observables are the $t$--dependences of the 
differential cross sections for various channels ($J/\psi, \phi ,
\rho, \pi$) at $|t| \ll 0.1\, \text{GeV}^2$, and their change with $x$;
at sufficiently large $Q^2$ such measurements can be related directly 
to the $t$--dependence of the gluon and quark GPDs at small $t$.
In particular, such measurements should be able to resolve
variations of the $t$--slope with $t$ and possible deviations from 
exponential $t$--dependence. Measurements of exclusive
processes require high luminosity and the capability to
detect the recoiling baryon at small angles, which is possible
with appropriate forward detectors. Another interesting option
are pion knockout processes, corresponding to exclusive scattering
from a pion at transverse distances $b \sim 1/M_\pi$, where both
the recoiling pion and the nucleon are identified in the final state;
see Ref.~\cite{Strikman:2003gz} for a detailed discussion.

The partonic content of the nucleon's pion cloud can in principle also 
be probed in high--energy $pp$ collisions with hard processes, such as 
dijet and Drell--Yan pair production. Such processes, including 
accompanying spectator interactions, are most naturally described 
in the transverse coordinate (impact parameter) representation employed
in our investigation here. Interesting new effects appear in collisions 
at multi--TeV energies (LHC), where the cross sections for hard processes 
can approach the geometric limit (black--disk regime) and the probability
for multiple hard interactions becomes significant. In this situation
it is important to realize that the $\pi B$ configurations participate
in the high--energy scattering process with a fixed transverse orientation,
which is frozen during the collision; depending on this orientation one 
may either have a violent collision of the pion with the other proton
or no interaction at all. The averaging over the orientations of the 
$\pi B$ configuration must be performed in the colliding $pp$ system 
with given transverse geometry, not in the partonic wave functions
of the individual protons. This circumstance affects 
\textit{e.g.} the rate of multijet events in peripheral 
collisions \cite{Rogers:2008ua}. More generally, the pion cloud 
represents an example of
transverse correlations in the nucleon's partonic wave function,
which are neglected in the usual mean--field approximation for 
high--energy $pp$ collisions. In particular, such correlations
play a role in central inclusive diffraction, where they reduce the 
rapidity gap survival probability relative 
to the mean--field result \cite{Frankfurt:2006jp}.
\acknowledgments
The authors are indebted to A.~Freund, J.~Goity, V.~Guzey, 
N.~Kivel, P.~Nadolsky, M.~V.~Polyakov, and A.~W.~Thomas for 
enlightening discussions and useful hints.

Notice: Authored by Jefferson Science Associates, LLC under U.S.\ DOE
Contract No.~DE-AC05-06OR23177. The U.S.\ Government retains a
non--exclusive, paid--up, irrevocable, world--wide license to publish or
reproduce this manuscript for U.S.\ Government purposes.
Supported by other DOE contracts.
\newpage
\appendix
\section{Meson--baryon couplings from $SU(3)$ symmetry}
\label{app:su3}
In this appendix we summarize the meson--baryon couplings used in 
the calculation of $SU(3)$ octet meson ($\pi, K, \eta$) contributions
to the sea quark distributions (see Sec.~\ref{subsec:sbar}) and derive 
the expressions for the corresponding meson momentum distributions 
in the nucleon.

For the coupling constants governing the $N \leftrightarrow M + B$ 
transitions we rely on $SU(3)$ flavor symmetry, which is known
to describe the empirical couplings well; see Ref.~\cite{Guzey:2005vz}
for a recent review. For $\bm{8} \leftrightarrow \bm{8} \times \bm{8}$ 
transitions there are two independent $SU(3)$--invariant structures,
with coupling constants traditionally denoted by $F$ and $D$.
With the standard assignments of the meson and baryon fields the 
Lagrangian takes the form (we show explicitly only the terms 
describing transitions $M + B \rightarrow p$)
\be
\mathcal{L} &=& 
g_8 \; \bar p \, \left[ \pi^+ n + {\textstyle\frac{1}{\sqrt{2}}} 
\pi^0 p 
\; - \; {\textstyle\frac{1}{\sqrt{6}}} (1 + 2 \alpha ) K^+ \Lambda
\right. \nonumber \\
&+& (1 - 2 \alpha ) 
\left( {\textstyle\frac{1}{\sqrt{2}}} K^+ \Sigma^0 
\; + \; K^0 \Sigma^+ \right) 
\nonumber \\
&-& \left. {\textstyle\frac{1}{\sqrt{6}}} (1 - 4 \alpha) \eta p
\right] \;\; + \;\; \textrm{h.c.}
\label{L_8_res}
\ee
Here $g_8 \equiv F + D$ is the overall octet coupling, which is 
related to our $\pi NN$ coupling as
\beq
g_{\pi NN} \;\; \equiv \;\; g_{\pi^0 p p} \;\; = \;\; 
g_8/\sqrt{2} ;
\eeq
we use $g_{\pi NN} = 13.05$ in our numerical calculations \cite{Koepf:1995yh}.
The ratio $\alpha \equiv F/(F + D)$ remains a free parameter in the 
context of $SU(3)$ flavor symmetry and can only be determined 
empirically or by invoking dynamical models. The $SU(6)$ spin--flavor 
symmetry of the non-relativistic quark model implies
$F/D = 2/3$, and thus 
\beq
\alpha \;\; = \;\; 2/5 \;\; = \;\; 0.4 ; 
\eeq
we use this value in our numerical studies in Sec.~\ref{subsec:sbar}.

For $\bm{8} \leftrightarrow \bm{8} \times \bm{10}$ transitions
there is only a single $SU(3)$--invariant structure, and the
Lagrangian is of the form
\be
\mathcal{L} &=& g_{10} \; \bar p 
\left( {\textstyle\frac{1}{\sqrt{3}}} \pi^+ \Delta^0 
+ {\textstyle\sqrt{\frac{2}{3}}} \pi^0 \Delta^+
\; - \; \pi^- \Delta^{++}
\right. 
\nonumber \\
&+& \left. {\textstyle\frac{1}{\sqrt{6}}} K^+ \Sigma^{\ast 0}
\; - \; {\textstyle\frac{1}{\sqrt{3}}} K^0 \Sigma^{\ast +} \right)
\;\; + \;\; \textrm{h.c.}
\label{L_10_res}
\ee
The decuplet coupling $g_{10}$ coincides (up to the sign) 
with our $\pi N \Delta$ coupling,
\beq
g_{\pi N \Delta} \;\; \equiv \;\; g_{\pi^- p \Delta^{++}} 
\;\; = \;\; -g_{10} ;
\eeq
we use $g_{\pi N \Delta} = 20.22$, which is close to the large--$N_c$ 
value of $(3/2) \, g_{\pi NN}$, \textit{cf.}\ Eq.~(\ref{g_largenc}).
Note that our definition of the coupling constant 
$g_{\pi N \Delta}$ differs from the one of Ref.~\cite{Koepf:1995yh}
by a factor, $g_{\pi N \Delta} (\text{Ref.~\cite{Koepf:1995yh}})  
= \sqrt{2} \, g_{\pi N \Delta} (\text{this work})$.

The GPDs of $SU(3)$ octet meson in the nucleon, $H_{MB}(y, t)$, 
and the corresponding impact parameter--dependent distributions 
are obtained by straightforward extension 
of the expressions for $\pi N$ and $\pi\Delta$ in Sec.~\ref{sec:chiral},
\textit{cf.}\ Eqs.~(\ref{H_pi_N_from_I}) and (\ref{H_pi_Delta_from_I}),
and Eqs.~(\ref{I_8_10})--(\ref{phi_10}). We work with the isoscalar 
distributions, in which we sum over the isospin components of the 
intermediate meson--baryon system. Using the couplings provided by 
Eqs.~(\ref{L_8_res}) and (\ref{L_10_res}), they are obtained as
(we omit the arguments for brevity)
\beq
\begin{array}{lclcl}
H_{\pi N} &\equiv& H_{\pi^+ n} + H_{\pi^0 p} &=&
3 \, g_{\pi NN}^2 \; I_8 , \\[1ex]
H_{\eta N} &\equiv& H_{\eta p} &=&
{\textstyle\frac{1}{3}} \, g_{\pi NN}^2 \, (1 - 4 \alpha)^2 \; I_8 , \\[1ex]
H_{K\Lambda} &\equiv& H_{K^+ \Lambda} &=&
{\textstyle\frac{1}{3}} \, g_{\pi NN}^2 \, (1 + 2 \alpha)^2 \; I_8 , \\[1ex]
H_{K\Sigma}  &\equiv& H_{K^+ \Sigma^0}  + H_{K^0 \Sigma^+} 
&=& 3 \, g_{\pi NN}^2 \, (1 - 2 \alpha)^2 \; I_8 , \\[1ex]
H_{\pi\Delta}  &\equiv& H_{\pi^+ \Delta^0} 
+ H_{\pi^0 \Delta^+} && \\[1ex]
 && + H_{\pi^- \Delta^{++}} &=& 2 \, g_{\pi N\Delta}^2 \; I_{10} , \\[1ex]
H_{K\Sigma^\ast}  &\equiv& H_{K^+ \Sigma^{\ast 0}} 
+ H_{K^0 \Sigma^{\ast +}} 
&=& {\textstyle\frac{1}{3}} \, g_{\pi N\Delta}^2 \; I_{10} .
\end{array}
\label{H_MB_summary}
\eeq
Here $I_8$ and $I_{10}$ are the basic momentum integrals of
Eqs.~(\ref{I_8_10})--(\ref{phi_10}), taken at the appropriate values of 
the meson and baryon masses. Note that because of isospin symmetry the 
distributions for the individual isospin components are all proportional 
to the same function and can be expressed in terms of the isoscalar 
distribution as
\beq
H_{\pi^+ n} \; = \; {\textstyle\frac{2}{3}} H_{\pi N},
\hspace{2em}
H_{\pi^0 p} \; = \; {\textstyle\frac{1}{3}} H_{\pi N}, 
\hspace{2em} \text{etc.}
\label{H_isospin}
\eeq
where the proportions are determined by the squares of the
coupling constants in the Lagrangian. The corresponding
$b$--dependent distributions, $f_{MB}(y, b)$, are then obtained by 
substituting these GPDs in Eq.~(\ref{f_pi_fourier}). 

To determine the coefficients with which the different mesons 
contribute to a given parton density in the proton, one must
account for the probability with which the parton occurs in 
the individual meson charge states. For example (in abbreviated
notation)
\be
(\bar d - \bar u)_p &=& f_{\pi^+ n} \; (\bar d - \bar u)_{\pi^+} 
+ f_{\pi^0 p} \; (\bar d - \bar u)_{\pi^0} 
\nonumber \\
&=& {\textstyle\frac{2}{3}} f_{\pi N} \; q^{\text{val}}_{\pi} ,
\ee
where $q^{\text{val}}_{\pi}$ denotes the valence quark distribution
in the pion, Eq.~(\ref{q_pi_val}), and we have used Eq.~(\ref{H_isospin}) 
for the impact parameter distribution and the isospin and charge
conjugation relations for the parton densities in the pion, 
$\bar u_{\pi^+} = d_{\pi^+}, \bar u_{\pi^0} = \bar d_{\pi^0}$.
\section{Evaluation of coordinate--space distributions}
\label{app:evaluation}
In this appendix we present expressions suitable for
numerical evaluation of the transverse coordinate--dependent 
distribution of pions in the nucleon and their partial radial integrals.
The $b$--dependent distribution, defined by Eq.~(\ref{f_pi_fourier}),
is evaluated as (we omit the argument $y$ and the subscript for brevity)
\be
f(b) &=& \int\frac{d^2 \Delta_\perp}{(2\pi )^2} \; 
e^{-i (\bm{\Delta}_\perp \bm{b})} \;  H(t) 
\label{f_fourier} \\
&=& \frac{1}{2\pi} \; \int_0^\infty d\Delta_\perp \; \Delta_\perp \;
J_0 (\Delta_\perp b) \;  H(t) 
\label{f_bessel}
\ee
where $b = |\bm{b}|, \; \Delta_\perp = |\bm{\Delta}_\perp|, \;
t = -\Delta_\perp^2$, and $J_0$ denotes the Bessel function. 
Equation~(\ref{f_bessel}) can be used to calculate the $f(b)$ 
corresponding to a numerically given $H(t)$, as obtained from 
evaluating the loop integral Eq.~(\ref{I_8_10}) with $\pi N$ form factors. 
In practice, since $H(t)$ shows only a power--like fall--off at 
large $-t > 0$, we multiply the integrand in Eq.~(\ref{f_bessel}) 
by an exponential convergence factor, 
$\exp (\epsilon t)$, calculate the integral numerically for finite $\epsilon$, 
and estimate the limiting value for $\epsilon \rightarrow 0$ 
from the numerical data at finite $\epsilon$.

From Eq.~(\ref{f_bessel}) we can also derive expressions for the
partial radial integrals of $f(b)$, including those with a weighting
factor $b^2$. Using standard identities for integrals of the 
Bessel function multiplied by powers of its argument,
we obtain for the integrals over the region $b < b_0$
\be
\int d^2 b \; \Theta (b < b_0) \, f(b)
\; =\;  b_0 \int_0^\infty \! d\Delta_\perp \, J_1 (z_0) \,  H(t)
\label{f_int_bessel}
\ee
\be
\int d^2 b \; \Theta (b < b_0) \; f(b) \; b^2 
\; = \; b_0^2 \int_0^\infty \frac{d\Delta_\perp}{\Delta_\perp}
\left[ 2 J_0 (z_0) \right. &&
\nonumber \\
+ \left. 
(z_0 - 4/z_0) \; J_1 (z_0) \right]\,  H(t) , \;\;\; &&
\label{f_int_b2_bessel}
\ee
where
\beq
z_0 \;\; \equiv \;\; \Delta_\perp b_0 .
\eeq
The complementary integrals over the region $b > b_0$ are 
calculated by re-writing the original Fourier integral 
for $f(b)$, Eq.~(\ref{f_fourier}), in the form
\beq
\int d^2 b \; \Theta (b > b_0) \;\; = \;\;
\int\limits d^2 b \;\; - \;\; \int d^2 b \; \Theta (b < b_0) .
\eeq
The unrestricted integral on the R.H.S.\ then produces a 
two--dimensional delta function at $\bm{\Delta}_\perp = 0$
and can be evaluated in terms of $H(t = 0)$ or its derivative.
In this way we obtain
\be
\lefteqn{\int d^2 b \; \Theta (b > b_0)  \; f(b)} && 
\nonumber \\ 
&=& H(t = 0) \; - \; \int d^2 b \; \Theta (b < b_0) \; f(b) , \;\;
\\
\lefteqn{
\int d^2 b \; \; \Theta (b > b_0) \; b^2 \; f(b)} && 
\nonumber \\
&=& 4 \frac{\partial H}{\partial t} (t = 0) \; - \; 
\int d^2 b \; \; \Theta (b < b_0) \; b^2 \; f(b) , \phantom{xx}
\ee
where the right--hand side can be evaluated using
Eqs.~(\ref{f_int_bessel}) and (\ref{f_int_b2_bessel}). 
Note that the $\Delta_\perp$ integrals representing the $b$--integrated
distributions, Eqs.~(\ref{f_int_bessel}) and (\ref{f_int_b2_bessel}), 
converge more rapidly at large $\Delta_\perp$ than the integral
representing the original $f(b)$, Eq.~(\ref{f_bessel}), and
can therefore more easily be computed numerically.
\end{document}